\documentclass[12pt,a4paper,english]{article}
\usepackage[T1]{fontenc}
\usepackage{babel}
\usepackage{amsmath}
\usepackage{amsfonts}
\usepackage{amsthm}
\usepackage{mathrsfs}

\makeatletter
 \newcommand{\lyxaddress}[1]{
   \par {\raggedright #1 
   \vspace{1.4em}
   \noindent\par}
 }
\makeatother

\makeatletter

\newcommand\appsection{\@startsection {section}{1}{\z@}%
{-3.5ex \@plus -1ex \@minus -.2ex}%
{2.3ex \@plus.2ex}%
{\normalfont\Large\bfseries\noindent%
Appendix.\hspace{0.5em}}}

\newtheorem{athm}{Theorem}[section]
\newtheorem{alem}[athm]{Lemma}
\newtheorem{aprop}[athm]{Proposition}
\newtheorem{acor}[athm]{Corollary}
\newtheorem{adef}[athm]{Definition}

\makeatother

\newcommand{\R}{\mathbb{R}}
\newcommand{\C}{\mathbb{C}}
\newcommand{\Z}{\mathbb{Z}}
\newcommand{\N}{\mathbb{N}}
\newcommand{\I}{\mathbb{I}}

\newcommand{\HH}{\mathcal{H}}
\newcommand{\KK}{\mathcal{K}}
\newcommand{\XX}{\mathcal{X}}
\newcommand{\BB}{\mathcal{B}}
\newcommand{\TT}{\mathcal{T}}

\newcommand{\gH}{\mathfrak{H}}
\newcommand{\gY}{\mathfrak{Y}}
\newcommand{\gU}{\mathfrak{U}}
\newcommand{\gC}{\mathfrak{C}}
\newcommand{\gT}{\mathfrak{T}}
\newcommand{\gV}{\mathfrak{V}}

\newcommand{\sD}{\mathscr{D}}

\DeclareMathOperator{\Dom}{Dom}
\DeclareMathOperator{\sgn}{sgn}
\DeclareMathOperator{\Span}{span}
\DeclareMathOperator{\cotgh}{cotgh}
\DeclareMathOperator{\Ker}{Ker}
\let\Im\undefined
\DeclareMathOperator{\Im}{Im}
\DeclareMathOperator{\supp}{supp}

\newtheorem{thm}{Theorem}
\newtheorem{lem}[thm]{Lemma}
\newtheorem{prop}[thm]{Proposition}

\theoremstyle{definition}
\newtheorem*{def*}{Definition}
\theoremstyle{remark}
\newtheorem*{rem*}{Remark}

\newcommand{\AC}{\widetilde{AC}}
\newcommand{\dd}{\textrm{d}}

\begin{document}

\title{Propagators 
weakly associated to a family of Hamiltonians and the adiabatic theorem for the Landau Hamiltonian with a time-dependent Aharonov-Bohm flux}

\author{J.~Asch$^{1,2}$, I.~Hradeck\'y$^3$,
  P.~\v{S}\v{t}ov\'{i}\v{c}ek$^3$}

\date{}

\maketitle

\lyxaddress{$^1$Centre de Physique Th\'eorique, CNRS, Luminy, Case
  907, Marseille Cedex 9, France }

\lyxaddress{$^2$CPT--PhyMat, Universit\'e du Sud Toulon--Var, BP 20132,
  F-83957 La Garde Cedex, France}

\lyxaddress{$^3$Department of Mathematics, Faculty of Nuclear Science,
  Czech Technical University, Trojanova 13, 120 00 Prague, Czech
  Republic}

\begin{abstract}
  We study the dynamics of a quantum particle moving in a plane under
  the influence of a constant magnetic field and driven by a slowly
  time-dependent singular flux tube through a puncture. The known  adiabatic results do not cover these models as the Hamiltonian  has  time dependent domain. We give a meaning to 
 the propagator and prove  an adiabatic theorem. To this end we introduce and develop the new notion of a propagator weakly associated to a time-dependent Hamiltonian.
\end{abstract}

\section{Introduction}

The model under consideration originates from Laughlin's
\cite{Laughlin} and Halperin's \cite{Halperin} discussion of the
Integer Quantum Hall effect. In the mathematical physics literature
Bellissard \cite{Bellissard} and Avron, Seiler, Simon
\cite{AvronSeilerSimon} used an adiabatic limit of the model (with
additional randomness) to introduce indices. The indices explain the
quantization of charge transport observed in the experiments
\cite{Klitzing}.

In this paper we discuss some mathematical aspects of the existence of
the propagator and the validity of the adiabatic approximation and
propose how to overcome the difficulties originating from the strong
singularity of the external field.

Let us specify the model, summarize our results and introduce the
notation. The configuration space is $\R^{2}\setminus\{(0,0)\}$ and
the model is considered in polar coordinates $(r,\theta)$. The vector
potential $A$ is the sum of a part for the homogeneous magnetic field
of strength $B>0$,
\[
\frac{B}{2}\,(x_{1}\dd x_{2}-x_{2}\dd x_{1})
= \frac{Br^{2}}{2}\,\dd\theta,
\]
plus a part describing the flux $\Phi$ which varies in time,
\[
\frac{\Phi}{2\pi}\,\frac{1}{\vert\vec{x}\vert^{2}}\,
(x_{1}\dd x_{2}-x_{2}\dd x_{1})
= \frac{\Phi}{2\pi} \,\dd\theta ;
\]
the real-valued function $\Phi$ is assumed to be monotonous and $C^2$.
With the metric coefficients $g_{11}=1$, $g_{22}=r^{2}$, $g_{12}=0$,
the differential expression of the Hamiltonian acting in
$L^{2}(\R_+\times\lbrack0,2\pi[\,,r\dd{}r\dd\theta)$ is
\begin{eqnarray*}
  && \frac{1}{2m}\left(-i\hbar\partial_{j}
    -\frac{e}{c}A_{j}\right)\sqrt{g}g^{jk}
  \left(-i\hbar\partial_{k}-\frac{e}{c}A_{k}\right) \\
  && =\frac{\hbar^{2}}{2m}
  \left(-\frac{1}{r}\partial_{r}r\partial_{r}+\frac{1}{r^{2}}
    \left(-i\partial_{\theta}-\frac{e}{\hbar c}\frac{Br^{2}}{2}
      -\frac{e}{h c}\,\Phi\right)^{\!2}\right).
\end{eqnarray*}
Our purpose is to study the response of the system if flux quanta
$hc/e$ are added adiabatically, i.e. the flux function is of the form
$t\mapsto\Phi(t/\tau)$ with the time $t$ varying in $[\,0,\tau\,]$ for
some $\tau\gg1$.

In a first step we analyze the case when $\Phi$ is linear.
Furthermore, we fix an angular momentum sector defined by
$-i\partial_{\theta}e^{im\theta}=me^{im\theta}$ $(m\in\Z)$, and use a
slow time $s$, i.e.: the substitution $s=-m+e/(h{}c)\Phi(t/\tau)$.
Also we are not interested here in keeping track of the behavior in
the physical parameters $e$, $\hbar$, $c$, $2m$, so we set them all
equal to one. This is our motivation to consider the operator
\begin{equation}
  \label{eq:defH}
  H(s) = -\frac{1}{r}\partial_{r}r\partial_{r}
  + \frac{1}{r^{2}}\left(s+\frac{Br^{2}}{2}\right)^{\!2}
  \quad\textrm{in }L^{2}(\R_+,r\dd r).
\end{equation}

In a second step we shall then show that our analysis generalizes to
Hamiltonians of the form $H\big(\zeta(s)\big)$ where $\zeta\in{}C^2$
is a monotone function.

$H(s)$ is essentially selfadjoint on $C_{0}^{\infty}(]0,\infty[)$ iff
$s^{2}\ge1$ \cite{ReedSimon2}. For $0<s^{2}<1$ we impose the regular
boundary condition as $r\to0+$ (i.e.: a wavefunction belongs to the
domain if it has no part proportional to the (square integrable)
singularity $r^{-|s|}$). This is in fact the most common choice, see
\cite{ExnerStovicekVytras} for a detailed discussion. The case $s=0$
is particular since the singularity in question is logarithmic but
otherwise the situation is similar, see \cite{Albeverioatal}. The
Hamiltonian $H(s)$ is unambiguously determined by specifying a
complete set of eigenfunctions with corresponding eigenvalues, see
below.

The dynamics of the model should be defined by
\begin{equation}
  \label{eq:Sch_eq_Utau}
  i\partial_{s}U_{\tau}(s,s_{0})\psi
  = \tau H(s) U_{\tau}(s,s_{0})\psi,
  \quad U_{\tau}(s_{0},s_{0})\psi=\psi,
\end{equation}
where $U_{\tau}$ is unitary and $\psi$ is an arbitrary initial
condition from the domain of $H(s_{0})$. The existence of a propagator
in this sense is, however, uncertain. The problem arises from the fact
that the domain of $H(s)$ is not constant in $s$, respectively that
$\dot H(s)$ is not relatively bounded with respect to $H(s)$. Thus the
usual theorems which assure the existence of the propagator
\cite{ReedSimon2} and the validity of the adiabatic approximation
\cite{AvronSeilerYaffe, AvronElgart} are not directly applicable.

A convenient way to see this is to consider the eigenfunctions.  The
operator $H(s)$ has a simple discrete spectrum; the eigenvalues are
\begin{equation}
  \label{eq:lambda_n}
  \lambda_{n}(s)=B(s+|s|+2n+1),\textrm{~}\quad n\in\{0,1,2,\ldots\},
\end{equation}
 with the corresponding normalized
eigenfunctions
\[
\varphi_{n}(s;r)=c_{n}(s)\, r^{|s|}\,
L_{n}^{(|s|)}\!\left(\frac{Br^{2}}{2}\right)
\exp\!\left(-\frac{Br^{2}}{4}\right)
\]
where
\[
c_{n}(s)=\left(\frac{B}{2}\right)^{\!(|s|+1)/2}\left(\frac{2\,
    n!}{\Gamma(n+|s|+1)}\right)^{\!1/2}
\]
are the normalization constants and $L_{n}^{(|s|)}$ are the
generalized Laguerre polynomials (see, for example,
\cite{ExnerStovicekVytras}).

The derivative of $H(s)$ equals
\[
\dot{H}(s)=\frac{2s}{r^{2}}+B.
\]
Notice that if $|s|\leq1$ then $\varphi_{n}(s)$ cannot belong to the
domain $\Dom\dot{H}(s)$ since $\dot{H}(s)\varphi_{n}(s)\sim
r^{-2+|s|}$ for $r\to0+$. This means that $\dot{H}(s)$ is not
relatively bounded with respect to $H(s)$.

Remark that, on the other hand, the quadratic expression
\[
\int_{0}^{\infty}\varphi_{m}(s;r)\,\dot{H}(s)\varphi_{n}(s;r)\,
r\mathrm{\,d}r
\]
makes good sense.  In order to avoid a complicated notation we shall
denote it by the symbol
$\langle\varphi_{m}(s),\dot{H}(s)\varphi_{n}(s)\rangle$ even though
the symbol cannot be taken literally and is therefore somewhat
misleading. Furthermore, the derivative of the eigenfunction,
$\dot{\varphi}_{n}(s)$, belongs to $L^{2}(\R_{+},r\,\textrm{d}r)$.
Since the eigenfunctions are chosen to be real-valued it holds true
that
\[
\left\langle \varphi_{n}(s),\dot{\varphi}_{n}(s)\right\rangle =0.
\]

Let us also note that, similarly, if $|s|\leq1$ and $s^{2}\neq s'^{2}$
then the eigenfunction $\varphi_{n}(s)$ cannot belong to $\Dom H(s')$.
It is so because (as formal expressions)
$H(s')-H(s)=(s'^{2}-s^{2})/r^{2}+B(s'-s)$ and $H(s')\varphi_{n}(s;r)$
has a non-integrable singularity at $r=0$. Hence $\Dom H(s)$ depends
on $s$.

It turns out that, following the strategy of Born and Fock
\cite{BornFock}, the problems of existence and adiabatic approximation
can both be handled:

denote the eigenprojector onto $\C\varphi_{n}(s)$ by $P_{n}(s)$; it is
differentiable as a bounded operator. The hard part of our work
consists in showing that
\[
i\sum_{k=0}^{\infty}\dot P_{k}(s)P_{k}(s)
\]
is a bounded operator. This is stated in Lemma \ref{thm:estimQs}. It
requires work because its matrix elements have bad off--diagonal
decay, see Lemma \ref{thm:bound-Qs} (which is formulated for the
unitarily equivalent operator $Q$).

Now 
\[
H_{AD}(s) := H(s)+\frac{i}{\tau}
\sum_{n=0}^{\infty}\dot P_{n}(s)P_{n}(s)
\]
has a propagator which is well defined in the usual way, i.e.
\begin{equation}
  \label{eq:defUAD}
  i\partial_{s} U_{AD}(s,s_{0})\psi=\tau H_{AD}(s)U_{AD}(s,s_{0})\psi,
  \quad U_{AD}(s_{0},s_{0})\psi=\psi,
\end{equation}
for $\psi\in\Dom\left(H_{AD}(s_{0})\right)$. To see this notice that
$U_{AD}$ can be computed by its action on the eigenbasis:
\[
U_{AD}(s,s_{0})\varphi_{n}(s_0)
= e^{-i\tau\int_{s_{0}}^{s}\lambda_{n}(u)\,\dd u}\varphi_{n}(s).
\]
Furthermore, $\lambda_{n}(s)-\lambda_{n}(0)$ is bounded in $n$ and
so 
$U_{AD}(s,s_{0})\Dom H_{AD}(s_{0})=\Dom{}H_{AD}(s)$.

Since $H(s)-H_{AD}(s)$ is bounded the domains of $H(s)$ and
$H_{AD}(s)$ are identical. By time-dependent transformation a natural
candidate for the propagator of $H(s)$ is
\begin{equation}
  \label{eq:UtauUAC}
  U_{\tau}(s,s_{0}) := U_{AD}(s,0)C(s,s_{0})U_{AD}(0,s_{0})
\end{equation}
where $C(s,s_0)$ is defined by
\begin{equation}
  \label{eq:defC}
  i\partial_{s}C(s,s_{0}) = -Q_{\tau}(s)C(s,s_{0}),\quad
  C(s_{0},s_{0}) = \I,
\end{equation}
with
\begin{equation}
  \label{eq:defQtau}
  Q_{\tau}(s) := U_{AD}(0,s)
  \left(i\sum_{k=0}^{\infty}\dot P_{k}(s)P_{k}(s)\right)U_{AD}(s,0).
\end{equation}
Since $\|Q_\tau(s)\|$ is locally bounded the propagator $C(s,s_0)$ is
well defined by the Dyson formula.

The adiabatic approximation problem is settled in Proposition
\ref{prop:AdiabApprox} were it is shown that
\[
\Vert U_{\tau}(s,0)-U_{AD}(s,0)\Vert
= O\!\left(\frac{1}{\tau}\right).
\]

It remains unclear, however, whether $C(s,s_{0})$ preserves the domain
of $H(0)$ and therefore whether the propagator $U_{\tau}(s,s_0)$ is
actually related to the Hamiltonian $H(s)$ in the usual sense. To
handle this problem we develop the general concept of weak association
of a propagator and a time dependent Hamiltonian.  We can show that
$U_{\tau}$ is weakly associated to $H(s)$ and that the Schrödinger
equation (\ref{eq:Sch_eq_Utau}) is fulfilled in the sense of
distributions.

We shall use the following notation. The symbol $V(s)$ stands for the
unitary operator which sends all eigenstates at time $0$ to the
corresponding eigenstates at time $s$, i.e.
\begin{equation}
  \label{eq:defV}
 \textrm{ }V(s)\varphi_{n}(0) = \varphi_{n}(s)\qquad
 \forall n\in\Z_+
\end{equation}
(here and everywhere in what follows $\Z_+$ stands for the set of
nonnegative integers). Further set
\begin{equation}
  W(s)=V(s)^{-1}H(s)V(s)=\sum_{n=0}^{\infty}\lambda_{n}(s)\,
  P_{n}(0)\label{eq:defW}
\end{equation}
and
\begin{equation}
  \Omega(s)=\sum_{n=0}^{\infty}\omega_{n}(s)\,
  P_{n}(0)\label{eq:defOmega}
\end{equation}
where
\[
\omega_{n}(s)=\int_{0}^{s}\lambda_{n}(u)\,\textrm{d}u.
\]

Remark  that the adiabatic propagator decomposes as
\[
U_{AD}(s,s_0)
= V(s)e^{-i\tau\left(\Omega(s)-\Omega(s_{0})\right)}V(s_{0})^{-1}.
\]

The paper is organized as follows. In
Sections~\ref{sec:auxiliary-estimates} and \ref{sec:boundedness_Q} we
do the analysis necessary to prove the boundedness result stated in
Lemma~\ref{thm:estimQs}. Section~\ref{sec:propagator} is devoted to
the existence problem for the propagator. In
Section~\ref{sec:adiabat_theorem} we prove the adiabatic theorem in
Proposition \ref{prop:AdiabApprox}. The result is then extended to a
more general time-dependence in Section~\ref{sec:general_dependence}.

A rather independent part of the paper is the Appendix where we
propose the notion of a propagator weakly associated to a
time-dependent Hamiltonian. We indicate cases where the weak
association can be verified while the usual relationship between a
propagator and a Hamiltonian is unclear or even is not valid. In
particular, this concept was inspired by the situation we encountered
in the present model. We believe, however, that this idea need not be
restricted to this case only and that it might turn out to be useful
in resolving this type of difficulties in other models as well.

\section{Auxiliary estimates of matrix operators}
\label{sec:auxiliary-estimates}

Here we derive some auxiliary estimates that will be useful later when
verifying assumptions of the adiabatic theorem.

\begin{lem}\label{thm:estim-movern}Let $A(\sigma)$ be an operator
  in $l^{2}(\N)$ depending on a parameter $\sigma\geq0$ whose matrix
  entries in the standard basis equal\[ A(\sigma)_{mn}=\left\{
    \begin{array}{ll}
      0 & \textrm{ for }m=n\\
      \noalign{\medskip}-\frac{i}{n}\left(\frac{m}{n}\right)^{\sigma} & \textrm{ for }m<n\\
      \noalign{\medskip}\frac{i}{m}\left(\frac{n}{m}\right)^{\sigma} &
      \textrm{ for }m>n\end{array}\right..\] Then $A(\sigma)$ is
  bounded, uniformly in $\sigma$, and its norm satisfies the
  estimate\[ \| A(\sigma)\|\leq24.\]
\end{lem}\begin{proof}The proof will be done in several steps.
  
  \emph{(i)} Let $K(\sigma)$ be an integral operator acting in
  $L^{2}(\R_{+},\textrm{d}x)$ with the integral kernel
  \[
  \mathcal{K}_{\sigma}(x,y) = \left\{ \begin{array}{ll}
      -\frac{i}{y}\,\big(\frac{x}{y}\big)^{\sigma} &
      \textrm{ for }x<y\\
      \noalign{\medskip}\frac{i}{x}\left(\frac{y}{x}\right)^{\sigma}
      & \textrm{ for }x>y\end{array}\right..
  \]
  Let us show that
  \[
  \| K(\sigma)\|=\frac{2}{2\sigma+1}\,.
  \]
  First we apply the unitary transform
  \begin{equation}
    \label{eq:unitaryU}
    U:L^{2}(\R_{+},\textrm{d}x)\to L^{2}(\R,\textrm{d}y),\textrm{~}
    U\psi(y) = e^{y/2}\psi(e^{y}).
  \end{equation}
  The inverse transform reads
  $U^{-1}\hat{\psi}(x)=x^{-1/2}\hat{\psi}(\ln x)$.  Set
  \[
  \tilde{K}(\sigma)=UK(\sigma)U^{-1}.
  \]
  One finds that $\tilde{K}(\sigma)$ is again an integral operator
  with the integral kernel
  \[
  \tilde{\mathcal{K}}_{\sigma}(y,z)=i\sgn(y-z)\,
  e^{-(\sigma+1/2)|y-z|}.
  \]
  Hence $\tilde{K}(\sigma)$ is a convolution operator and it is
  therefore diagonalizable with the aid of the Fourier transform
  $\mathcal{F}$ on $\R$. This means that
  \begin{displaymath}
    \big(\mathcal{F}\tilde{K}(\sigma)\mathcal{F}^{-1}\psi\big)(z)
    = \hat{q}(z)\psi(z)
  \end{displaymath}
  where
  \[
  \hat{q}(z) = \int_{\R}e^{izy}\sgn(y)\, 
  e^{-(\sigma+1/2)|y|}\,\textrm{d}y
  = \frac{2iz}{\left(\sigma+\frac{1}{2}\right)^{\!2}+z^{2}}\,.
  \]
  It follows that
  \begin{equation}
    \| K(\sigma)\|=\|\mathcal{F}\tilde{K}(\sigma)\mathcal{F}^{-1}\|
    =\|\hat{q}\|_{\infty}=\frac{1}{\sigma+\frac{1}{2}}\,.
    \label{eq:normKs-xoverys}
  \end{equation}
  
  \emph{(ii)} Suppose that $\{\psi\}_{n=1}^{\infty}$ is an orthogonal
  system in $L^{2}(\R_{+},\textrm{d}x)$ such that
  \[
  \forall m,n\in\N,\ \left\langle\psi_{m},K(\sigma)
    \psi_{n}\right\rangle = A(\sigma)_{mn}
  \]
  and
  \[
  \forall n\in\N,\ \|\psi_{n}\|^{2}=\kappa>0.
  \]
  Let $P_{+}$ be the orthogonal projector onto
  $\Span\{\psi_{n}\}_{n=1}^{\infty}$ in $L^{2}(\R_{+},\textrm{d}x)$.
  Then one can identify $P_{+}K(\sigma)P_{+}$ with
  $\kappa^{-1}A(\sigma)$. Hence
  \begin{equation}
    \label{eq:As_estim_Ks}
    \| A(\sigma)\|=\kappa\| P_{+}K(\sigma)P_{+}\|
    \leq\kappa\|K(\sigma)\|.
  \end{equation}
  
  \emph{(iii)} We shall construct an orthogonal system
  $\{\psi_{n}\}_{n=1}^{\infty}$ described in the preceding point as
  follows. Consider the natural embedding
  $L^{2}([n,n+1],\textrm{d}x)\subset L^{2}(\R_{+},\textrm{d}x)$,
  $n\in\N$. We seek $\psi_{n}\in L^{2}([n,n+1],\textrm{d}x)$ in the
  form
  \[
  \psi_{n}=\alpha_{n}u_{n}+\beta_{n}v_{n}+f_{n}
  \]
  where $\alpha_{n},\beta_{n}\in\R$,
  $u_{n},v_{n},f_{n}\in{}L^{2}([n,n+1],\textrm{d}x)$,
  \[
  u_{n}(x)=x^{\sigma},\  v_{n}(x)=x^{-\sigma-1}
  \textrm{~for~}x\in[n,n+1],
  \]
  and $f_{n}\perp u_{n}$, $f_{n}\perp v_{n}$. Suppose for definiteness
  that $m<n$. Then
  \begin{eqnarray*}
    \left\langle \psi_{m},K(\sigma)\psi_{n}\right\rangle 
    & = & \int_{m}^{m+1}\textrm{d}x\int_{n}^{n+1}\textrm{d}y\,
    \mathcal{K}_{\sigma}(x,y)\,\psi_{m}(x)\,\psi_{n}(y)\\
    & = & -i\left\langle u_{m},\psi_{m}\right\rangle
    \left\langle v_{n},\psi_{n}\right\rangle .
  \end{eqnarray*}
  Furthermore,
  \[
  \left\langle \psi_{n},K(\sigma)\psi_{n}\right\rangle
  =\int_{n}^{n+1}\int_{n}^{n+1}\mathcal{K}_{\sigma}(x,y)\,
  \psi_{n}(x)\,\psi_{n}(y)\,\textrm{d}x\textrm{d}y=0
  \]
  since $\mathcal{K}_{\sigma}(x,y)$ is antisymmetric,
  $\mathcal{K}_{\sigma}(y,x)=-\mathcal{K}_{\sigma}(x,y)$.
  Consequently, it suffices to choose the real coefficients
  $\alpha_{n}$, $\beta_{n}$ so that
  \[
  \forall n\in\N,\ \left\langle u_{n},\psi_{n}\right\rangle
  = n^{\sigma},\ \left\langle v_{n},\psi_{n}\right\rangle
  = n^{-\sigma-1}.
  \]
  This system has a unique solution $(\alpha_{n},\beta_{n})$. The
  function $f_{n}$ can be arbitrary. Its only purpose is to adjust the
  norms of the functions $\psi_{n}$ so that they are all equal. Set
  \[
  N_{n}(\sigma)=\|\alpha_{n}u_{n}+\beta_{n}v_{n}\|^{2}
  =\int_{n}^{n+1}\left(\alpha_{n}x^{\sigma}
    +\beta_{n}x^{-\sigma-1}\right)^{\!2}\textrm{d}x
  \]
  and
  \[
  \kappa(\sigma)=\sup_{n\in\N}N_{n}(\sigma).
  \]
  One can choose the orthogonal system $\{\psi_{n}\}_{n=1}^{\infty}$
  so that $\|\psi_{n}\|^{2}=\kappa(\sigma)$ for all $n$. According to
  (\ref{eq:normKs-xoverys}) and (\ref{eq:As_estim_Ks}) we have
  \begin{equation}
    \label{eq:Askappas}
    \| A(\sigma)\|\leq\frac{2\,\kappa(\sigma)}{2\sigma+1}\,.
  \end{equation}
  
  \emph{(iv)} It remains to find an upper bound on $\kappa(\sigma)$.
  Set
  \[
  \xi_{n}=n^{\sigma},\ \eta_{n}=n^{-\sigma-1}.
  \]
  Simple algebraic manipulations yield
  \[
  N_{n}(\sigma)=\frac{\left\langle v_{n},v_{n}\right\rangle
    \xi_{n}^{\,2}-2\left\langle u_{n},v_{n}\right\rangle
    \xi_{n}\eta_{n}+\left\langle u_{n},u_{n}\right\rangle
    \eta_{n}^{\,2}}{\left\langle u_{n},u_{n}\right\rangle
    \left\langle v_{n},v_{n}\right\rangle
    -\left\langle u_{n},v_{n}\right\rangle ^{2}}\,.
  \]
  Here
  \begin{eqnarray*}
    \left\langle u_{n},v_{n}\right\rangle
    & = & \ln\!\left(1+\frac{1}{n}\right),\\
    \left\langle u_{n},u_{n}\right\rangle
    & = & \frac{1}{2\sigma+1}
    \left((n+1)^{2\sigma+1}-n^{2\sigma+1}\right),\\
    \left\langle v_{n},v_{n}\right\rangle
    & = & \frac{1}{2\sigma+1}\left(n^{-2\sigma-1}
      -(n+1)^{-2\sigma-1}\right).
  \end{eqnarray*}
  Set
  \[
  w=\left(\sigma+\frac{1}{2}\right)\ln\!\left(1+\frac{1}{n}\right).
  \]
  One can rewrite the expression for $N_{n}(\sigma)$ as follows,
  \[
  N_{n}(\sigma)=\frac{2\sigma+1}{n}\,
  \frac{\sinh(w)\cosh(w)-w}{\sinh^{2}(w)-w^{2}}\,.
  \]
  Using an elementary analysis one can show that
  \[
  \frac{\sinh(w)\cosh(w)-w}{\sinh^{2}(w)-w^{2}}
  \leq\frac{\sinh(w)\cosh(w)-w}{\sinh(w)\left(\sinh(w)-w\right)}
  \leq4\cotgh(w).
  \]
  Hence
  \[
  N_{n}(\sigma)\leq\frac{4(2\sigma+1)}{n}\,
  \frac{\left(1+\frac{1}{n}\right)^{\!2\sigma+1}+1}
  {\left(1+\frac{1}{n}\right)^{\!2\sigma+1}-1}\leq12(2\sigma+1).
  \]
  Consequently,
  \begin{equation}
    \label{eq:kappas-upperb}
    \kappa(\sigma)\leq12(2\sigma+1).
  \end{equation}
  From (\ref{eq:Askappas}) and (\ref{eq:kappas-upperb}) it follows
  that $\|A(\sigma)\|\leq24$.
\end{proof}

\begin{lem}
  \label{thm:estim_fs}
  Let $A(\sigma)$ be an operator in $l^{2}(\N)$ whose matrix entries
  in the standard basis equal
  \[
  A(\sigma)_{mn}=\left\{
    \begin{array}{ll}
      0 & \textrm{ for }m=n\\
      \noalign{\medskip}-\frac{i}{n}\,
      f_{\sigma}\!\left(\frac{m}{n}\right) & \textrm{ for }m<n\\
      \noalign{\medskip}\frac{i}{m}\,
      f_{\sigma}\!\left(\frac{n}{m}\right) & \textrm{ for }
      m>n\end{array}\right.
  \]
  where
  \[
  f_{\sigma}(u)=\frac{1-u^{\sigma}}{1-u}\,,\ u\in\,]0,1[\,,
  \]
  and $\sigma\in[0,1]$ is a parameter. Then $A(\sigma)$ is bounded and
  its norm satisfies the estimate
  \[ \|A(\sigma)\|\leq\left(\frac{\sqrt{2}}{3}+4\right)\pi^{2}\sigma.
  \]
\end{lem}

\begin{proof}
  The proof will be done in several steps.
  
  \emph{(i)} Let $K(\sigma)$ be an integral operator acting in
  $L^{2}(\R_{+},\textrm{d}x)$ with the integral kernel
  \[
  \mathcal{K}_{\sigma}(x,y) = \left\{
    \begin{array}{ll}
      -\frac{i}{y}\, f_{\sigma}\big(\frac{x}{y}\big) & \textrm{ for }x<y\\
      \noalign{\medskip}\frac{i}{x}\,
      f_{\sigma}\!\left(\frac{y}{x}\right) & \textrm{ for }x>y
    \end{array}
  \right..
  \]
  Let us show that
  \begin{equation} 
    \label{eq:normKs}
    \|K(\sigma)\|\leq\pi^{2}\sigma.
  \end{equation}
  This step is quite analogous to the proof of point (i) in
  Lemma~\ref{thm:estim-movern}. First we apply the unitary transform
  $U$ defined in (\ref{eq:unitaryU}). Set
  \[
  \tilde{K}(\sigma)=UK(\sigma)U^{-1}.
  \]
  One finds that $\tilde{K}(\sigma)$ is again an integral operator
  with the integral kernel
  \[
  \tilde{\mathcal{K}}_{\sigma}(y,z)=i\sgn(y-z)\,
  f_{\sigma}\!\left(e^{-|y-z|}\right)e^{-|y-z|/2}.
  \]
  Thus $\tilde{K}(\sigma)$ is a convolution operator which is
  diagonalizable with the aid of the Fourier transform $\mathcal{F}$
  on $\R$. This means that
  $\big(\mathcal{F}\tilde{K}(\sigma)\mathcal{F}^{-1}\psi\big)(z)
  =\hat{q}(z)\psi(z)$
  where
  \[
  \hat{q}(z)=\int_{\R}e^{izy}\sgn(y)\,
  f_{\sigma}\!\left(e^{-|y|}\right)e^{-|y|/2}\,\textrm{d}y.
  \]
  A standard estimate yields
  \[
  |\hat{q}(z)|\leq2\int_{0}^{\infty}\frac{1-e^{-\sigma y}}{1-e^{-y}}\,
  e^{-y/2}\,\textrm{d}y\leq\sigma\int_{0}^{\infty}\frac{y}{\sinh(y/2)}\,
  \textrm{d}y=\pi^{2}\sigma.
  \]
  It follows that
  \[
  \|K(\sigma)\|=\|\mathcal{F}\tilde{K}(\sigma)\mathcal{F}^{-1}\|
  =\|\hat{q}\|_{\infty}\leq\pi^{2}\sigma.
  \]
  
  \emph{(ii)} Let $\chi_{n}(x)$ be the characteristic function of the
  interval $\,]n,n+1[\,$. The linear mapping
  \[
  J:l^{2}(\N)\to L^{2}(\R_{+},\textrm{d}x):
  \{\xi_{n}\}\mapsto\sum_{n=1}^{\infty}\xi_{n}\chi_{n}
  \]
  is an isometry. The adjoint mapping reads
  \[
  J^{*}:L^{2}(\R_{+},\textrm{d}x)\to
  l^{2}(\N):\psi\mapsto\{\left\langle \chi_{n},\psi\right\rangle
  \}_{n=1}^{\infty}.
  \]
  Set
  \[
  L(\sigma)=JA(\sigma)J^{*}.
  \]
  $L(\sigma)$ is an integral operator with the kernel
  \[
  \mathcal{L}_{\sigma}(x,y)=\sum_{m=1}^{\infty}
  \sum_{n=1}^{\infty}A(\sigma)_{mn}\chi_{m}(x)\chi_{n}(y).
  \]
  This can be rewritten as
  \[
  \mathcal{L}_{\sigma}(x,y)=\left\{
    \begin{array}{ll}
      -\frac{i}{[y]}\, f_{\sigma}\!\left(\frac{[x]}{[y]}\right) &
      \textrm{ if }0<[x]<[y]\\
      \noalign{\medskip}\frac{i}{[x]}\, f_{\sigma}\!
      \left(\frac{[y]}{[x]}\right) & \textrm{ if }0<[y]<[x]\\
      \noalign{\medskip}0 & \textrm{ otherwise}
    \end{array}
  \right..
  \]
  Here $[x]$ denotes the integer part of $x$. Notice that $J^{*}J$ is
  the identity on $l^{2}(\N)$ and so $L(\sigma)J=JA(\sigma)$.
  Consequently,
  \begin{equation}
    \label{eq:AsLs}
    \| A(\sigma)\|=\| JA(\sigma)\|
    =\|L(\sigma)J\|\leq\| L(\sigma)\|.
  \end{equation}
  
  \emph{(iii)} Denote by $\tilde{P}_{n}$, $n\in\Z_{+}$, the orthogonal
  projector onto $\C\chi_{n}$ in $L^{2}(\R_{+},\textrm{d}x)$. Set
  \[
  K^{\textrm{off}}(\sigma)=K(\sigma)-\tilde{P}_{0}K(\sigma)
  -K(\sigma)\tilde{P}_{0}+\tilde{P}_{0}K(\sigma)\tilde{P}_{0}
  -\sum_{n=1}^{\infty}\tilde{P}_{n}K(\sigma)\tilde{P}_{n}.
  \]
  In other words, we subtract from $K(\sigma)$ the diagonal as well as
  the first row and the first column (i.e., with index $0$) with
  respect to the orthogonal system $\{\chi_{n}\}_{n=0}^{\infty}$. We
  can say also that the integral kernel
  $\mathcal{K_{\sigma}^{\textrm{off}}}(x,y)$ vanishes if $[x]=[y]$ or
  $[x]=0$ or $[y]=0$ and otherwise it coincides with
  $\mathcal{K}_{\sigma}(x,y)$. Since
  \[
  \left\Vert\tilde{P}_{0}K(\sigma)\tilde{P}_{0}
    -\sum_{n=1}^{\infty}\tilde{P}_{n}K(\sigma)\tilde{P}_{n}\right\Vert
  =\sup_{n\in\Z_{+}}\|\tilde{P}_{n}K(\sigma)\tilde{P}_{n}\|
  \leq\|K(\sigma)\|
  \]
  we have
  \begin{equation}
    \label{eq:KoffK}
    \|K^{\textrm{off}}(\sigma)\|\leq4\|K(\sigma)\|.
  \end{equation}
  
  \emph{(iv)} It remains to estimate the norm of the difference
  $L(\sigma)-K^{\textrm{off}}(\sigma)$.  This is a Hermitian integral
  operator whose kernel does not vanish only if $0<[x]<[y]$ or
  $0<[y]<[x]$. Suppose for definiteness that $0<[x]<[y]$. Then the
  kernel equals, up to the multiplier $-i$,
  \begin{eqnarray*}
    \frac{1}{[y]}\, f_{\sigma}\!\left(\frac{[x]}{[y]}\right)
    -\frac{1}{y}\, f_{\sigma}\!\left(\frac{x}{y}\right)
    & = & \left(\frac{1}{[y]^{\sigma}}
      -\frac{1}{y^{\sigma}}\right)
    \frac{[y]^{\sigma}-[x]^{\sigma}}{[y]-[x]}\\
    & & +\,\frac{1}{y^{\sigma}}
    \left(\frac{[y]^{\sigma}-[x]^{\sigma}}{[y]-[x]}-
      \frac{y{}^{\sigma}-x{}^{\sigma}}{y-x}\right).
  \end{eqnarray*}
  Let us show that
  \begin{equation}
    \label{eq:fsfs01}
    0\leq\frac{1}{[y]}\,
    f_{\sigma}\!\left(\frac{[x]}{[y]}\right)-\frac{1}{y}\,
    f_{\sigma}\!\left(\frac{x}{y}\right)
    \leq\frac{2\sigma}{[x]([y]-[x])}.
  \end{equation}
  
  First notice that
  \[
  0\leq\frac{1}{[y]^{\sigma}}-\frac{1}{y^{\sigma}}
  =-\sigma\int_{y}^{[y]}z^{-\sigma-1}\textrm{d}z
  \leq\frac{\sigma(y-[y])}{[y]^{\sigma+1}}
  \]
  and so
  \begin{equation}
    \label{eq:1over_ys_diff}
    0\leq\left(\frac{1}{[y]^{\sigma}}-\frac{1}{y^{\sigma}}\right)
    \frac{[y]^{\sigma}-[x]^{\sigma}}{[y]-[x]}
    \leq\frac{\sigma}{[y]([y]-[x])}\,.
  \end{equation}
  Further set temporarily
  \begin{eqnarray*}
    D & = & \frac{[y]^{\sigma}-[x]^{\sigma}}{[y]-[x]}
    -\frac{y^{\sigma}-x^{\sigma}}{y-x}\\
    & = & \sigma\int_{0}^{1}
    \left(\left([x](1-t)+[y]t\right)^{\sigma-1}
      -\left(x(1-t)+yt\right)^{\sigma-1}\right)\textrm{d}t\,.
  \end{eqnarray*}
  The integrand in the last integral equals
  \[
  \sigma(1-\sigma)\xi_{t}^{\,\sigma-2}\left((x-[x])(1-t)+(y-[y])t\right)
  \]
  where $\xi_{t}$ is a real number lying between $[x](1-t)+[y]t$ and
  $x(1-t)+yt$. Notice that
  \[
  0\leq(x-[x])(1-t)+(y-[y])t\leq1.
  \]
  We assume that $0\leq\sigma\leq1$. Therefore
  \[
  0\leq D\leq\sigma(1-\sigma)\int_{0}^{1}
  \left([x](1-t)+[y]t\right)^{\sigma-2}\textrm{d}t
  =-\sigma\,\frac{[y]^{\sigma-1}-[x]^{\sigma-1}}{[y]-[x]}
  \]
  and so
  \begin{equation}
    0\leq\frac{1}{y^{\sigma}}D
    \leq\frac{\sigma[x]^{\sigma-1}}{y^{\sigma}([y]-[x])}
    \leq\frac{\sigma}{[x]([y]-[x])}\,.\label{eq:Dys01}
  \end{equation}
  Inequalities (\ref{eq:1over_ys_diff}) and (\ref{eq:Dys01}) jointly
  imply (\ref{eq:fsfs01}).
  
  \emph{(v)} From estimate (\ref{eq:fsfs01}) one can deduce that
  $L(\sigma)-K^{\textrm{off}}(\sigma)$ is a Hilbert-Schmidt operator
  and
  \begin{equation}
    \label{eq:LKoff01}
    \|L(\sigma)-K^{\textrm{off}}(\sigma)\|_{\textrm{HS}}
    \leq\frac{\sqrt{2}\,\pi^{2}}{3}\,\sigma.
  \end{equation}
  Actually,
  \begin{eqnarray*}
    \| L(\sigma)-K^{\textrm{off}}(\sigma)\|_{\textrm{HS}}^{\,2}
    & = & 2\int_{1}^{\infty}\textrm{d}x
    \int_{[x]+1}^{\infty}\textrm{d}y
    \left|\mathcal{L_{\sigma}}(x,y)
      -\mathcal{K_{\sigma}^{\textrm{off}}}(x,y)\right|^{2}\\
    & \leq & 8\sigma^{2}\int_{1}^{\infty}\textrm{d}x\,
    \frac{1}{[x]^{2}}\int_{[x]+1}^{\infty}\textrm{d}y\,
    \frac{1}{([y]-[x])^{2}}\\
    & = & 8\sigma^{2}\left(\sum_{k=1}^{\infty}
      \frac{1}{k^{2}}\right)^{\!2}.
  \end{eqnarray*}
  
  \emph{(vi)} Inequalities (\ref{eq:AsLs}), (\ref{eq:KoffK}),
  (\ref{eq:normKs}) and (\ref{eq:LKoff01}) imply that
  \[
  \|A(\sigma)\|\leq\| L(\sigma)\|\leq\| K^{\textrm{off}}(\sigma)\|
  +\|L(\sigma)-K^{\textrm{off}}(\sigma)\|
  \leq4\pi^{2}\sigma+\frac{\sqrt{2}\,\pi^{2}}{3}\,\sigma.
  \]
  This shows the lemma.
\end{proof}

\begin{lem}
  \label{thm:bound-As}
  Let $A(\sigma)$ be an operator in $l^{2}(\N)$ with the matrix entries
  in the standard basis
  \[
  A(\sigma)_{mn}=\left\{
    \begin{array}{ll}
      0 & \textrm{ for }m=n\\
      \noalign{\medskip}\frac{i}{n-m}\,
      \min\{\left(\frac{m}{n}\right)^{\sigma},
      \left(\frac{n}{m}\right)^{\sigma}\}
      & \textrm{ for }m\neq n
    \end{array}
  \right..
  \]
  Then $A(\sigma)$ is bounded for all $0\leq\sigma$ and its norm
  satisfies the estimate
  \[
  \|A(\sigma)\|\leq\pi+\left(\frac{\sqrt{2}}{3}+4\right)\pi^{2}\sigma.
  \]
\end{lem}

\begin{proof}
  Let us first show that
  \[
  \| A(0)\|\leq\pi.
  \]
  For $\sigma=0$ we get
  \[
  A(0)_{mn}=\frac{i}{n-m}\quad\textrm{if }m\neq n.
  \]
  Considering the natural embedding $l^{2}(\N)\subset l^{2}(\Z)$ let
  us denote by $P_{+}$ the orthogonal projector onto $l^{2}(\N)$ in
  $l^{2}(\Z)$. Let $B$ be an operator in $l^{2}(\Z)$ with the matrix
  \[
  B_{mn}=q(n-m)\textrm{ where }q(n)=\left\{ \begin{array}{ll}
      0 & \textrm{ for }n=0\\
      \noalign{\medskip}\frac{i}{n} &
      \textrm{~for~}n\neq0\end{array}\right..
  \]
  One can identify $A(0)$ with $P_{+}BP_{+}$. $B$ is a convolution
  operator and therefore it is diagonalizable by the Fourier transform
  $\mathcal{F}:l^{2}(\Z)\to L^{2}([0,2\pi],\textrm{d}\theta)$. In more
  detail,
  \[
  \left(\mathcal{F}B\mathcal{F}^{-1}\psi\right)\!(\theta)
  =\hat{q}(\theta)\psi(\theta)\textrm{ where }\hat{q}(\theta)
  =\sum_{n\in\Z}q(n)\, e^{in\theta}.
  \]
  One finds that $\hat{q}(\theta)=-\pi+\theta$. Consequently,
  \[
  \| A(0)\|=\| P_{+}BP_{+}\|\leq\| B\|
  =\|\mathcal{F}B\mathcal{F}^{-1}\|
  =\max_{\theta\in[0,2\pi]}|\hat{q}(\theta)|=\pi.
  \]

  Suppose now that $0<m<n$. Notice that
  \[
  \left(A(\sigma+1)-A(\sigma)\right)_{mn}
  =-\frac{i}{n}\,\left(\frac{m}{n}\right)^{\!\sigma}
  \]
  and
  \[
  \left(A(\sigma)-A(0)\right)_{mn}
  =-\frac{i}{n}\, f_{\sigma}\!\left(\frac{m}{n}\right).
  \]
  Using Lemma~\ref{thm:estim-movern} and Lemma~\ref{thm:estim_fs} one
  can estimate
  \begin{eqnarray*}
    \| A(\sigma)\| & \leq & \| A(0)\|+\| A(\sigma-[\sigma])-A(0)\|
    +\| A(\sigma-[\sigma]+1)-A(\sigma-[\sigma])\|\\
    &  & +\ldots+\,\| A(\sigma)-A(\sigma-1)\|\\
    & \leq & \pi+\left(\frac{\sqrt{2}}{3}+4\right)\pi^{2}
    (\sigma-[\sigma])+24[\sigma]\\
    & \leq & \pi+\left(\frac{\sqrt{2}}{3}+4\right)\pi^{2}\sigma.
  \end{eqnarray*}
  This proves the lemma.
\end{proof}

\section{Boundedness of the operator
  $i\sum_{k=0}^{\infty}\dot{P}_{k}(s)P_{k}(s)$}
\label{sec:boundedness_Q}

We consider $i\sum_{k=0}^{\infty}\dot{P}_{k}(s)P_{k}(s)$ in the time
independent frame, i.e. the operator $Q(s)$ defined by
\begin{equation}
  \label{eq:Qs_def}
  Q(s)=iV(s)^{\ast}\sum_{k=0}^{\infty}\dot{P}_{k}(s)P_{k}(s)V(s)
  =-i\dot{V}(s)^{\ast}V(s)=iV(s)^{\ast}\dot{V}(s).
\end{equation}
The operator $V(s)$ is defined in (\ref{eq:defV}). $Q(s)$ is symmetric
and its matrix entries in the basis $\{\varphi_{n}(0)\}$ are
\[
\left\langle \varphi_{m}(0),Q(s)\varphi_{n}(0)\right\rangle
=i\left\langle \varphi_{m}(s),\dot{\varphi}_{n}(s)\right\rangle .
\]
Since $\varphi_{n}(s)$ depends on $s$ only through the absolute value
it holds true that $Q(-s)=-Q(s)$ for $s\neq0$. For $s=0$ the
operator-valued function $Q(s)$ has a discontinuity. The goal of this
section is to show that the operator $Q(s)$ is in fact bounded.

To compute the matrix entries one can use the identity
\begin{equation}
  \left\langle \varphi_{m}(s),\dot{\varphi}_{n}(s)\right\rangle
  =\frac{\left\langle \varphi_{m}(s),
      \dot{H}(s)\varphi_{n}(s)\right\rangle }
  {\lambda_{n}(s)-\lambda_{m}(s)}\,.
  \label{eq:phi_phidot}
\end{equation}
Let us emphasize once more that the scalar product on the RHS should
be interpreted as a quadratic form since, in general,
$\varphi_{n}(s)\not\in\Dom\dot{H}(s)$.  The derivation goes through
basically as usual even though one cannot use the scalar product
directly. Differentiating the equation on eigenvalues one arrives at
the equality
\[
H(s)\dot{\varphi}_{n}(s;r)+\dot{H}(s)\varphi_{n}(s;r)
=\dot{\lambda}_{n}(s)\varphi_{n}(s;r)
+\lambda_{n}(s)\dot{\varphi}_{n}(s;r),
\]
valid for any $r>0$, in which one should substitute for $H(s)$ and
$\dot{H}(s)$ the corresponding formal differential operators. Next one
multiplies the equality by $r\varphi_{m}(s;r)$ and integrates the both
sides from $\varepsilon$ to infinity for some $\varepsilon>0$. In the
integral
\[
-\int_{\varepsilon}^{\infty}\varphi_{m}(s;r)
\partial_{r}r\partial_{r}\dot{\varphi}_{n}(s;r)\,\textrm{d}r
\]
occurring on the LHS side one integrates twice by parts. Checking the
asymptotic behavior of the eigenfunctions near the origin,
\begin{equation}
  \varphi_{n}(s;r)\sim\left(\frac{B}{2}\right)^{\!(|s|+1)/2}
  \left(\frac{2\, n!}{\Gamma(n+|s|+1)}\right)^{\!1/2}r^{|s|}
  \left(1+O\!\left(r^{2}\right)\right)\quad\textrm{for }r\to0+,
  \label{eq:asym-phi}
\end{equation}
one  finds that
\[
\lim_{r\to0+}r\varphi_{m}(s;r)\partial_{r}\dot{\varphi}_{n}(s;r)
=\lim_{r\to0+}r\left(\partial_{r}\varphi_{m}(s;r)\right)
\dot{\varphi}_{n}(s;r)=0.
\]
Hence sending $\varepsilon$ to $0$ actually leads to equality
(\ref{eq:phi_phidot}).

\begin{lem}\label{thm:bound-Qs}
  The matrix entries of the operator $Q(s)$ for $s\neq0$ are given by
  the formulae
  \[
  \left\langle
    \varphi_{m}(0),Q(s)\varphi_{n}(0)\right\rangle
  =0\quad\textrm{for~}m=n,
  \]
  and
  \[
  \left\langle
    \varphi_{m}(0),Q(s)\varphi_{n}(0)\right\rangle
  =\frac{i\sgn(s)}{2(n-m)}\,\min\left\{
    \frac{\gamma_{m}(s)}{\gamma_{n}(s)},
    \frac{\gamma_{n}(s)}{\gamma_{m}(s)}\right\}
  \quad\textrm{for }m\neq n,
  \]
  where
  \begin{equation}
    \label{eq:def_gamma}
    \gamma_{n}(s)=\left(\frac{\Gamma(n+|s|+1)}{n!}\right)^{\negmedspace1/2}.
  \end{equation}
\end{lem}

\begin{proof}
  Assume that $m<n$ and $s>0$. Using the explicit expression for the
  generalized Laguerre polynomials,
  \[
  L_{n}^{(\alpha)}(x)=\sum_{k=0}^{n}(-1)^{k}
  \binom{n+\alpha}{n-k}\frac{1}{k!}\,x^{k},
  \] one finds that
  \begin{eqnarray*}
    \left\langle \varphi_{m}(s),\dot{H}(s)\varphi_{n}(s)\right\rangle
    & = & 2s\, c_{m}(s)\, c_{n}(s)\\
    &  & \times\int_{0}^{\infty}r^{2s-1}L_{m}^{(s)}\!
    \left(\frac{Br^{2}}{2}\right)L_{n}^{(s)}\!
    \left(\frac{Br^{2}}{2}\right)
    \exp\!\left(-\frac{Br^{2}}{2}\right)\textrm{d}r\\
    & = & s\, c_{m}(s)\, c_{n}(s)\left(\frac{2}{B}\right)^{\! s}S_{m,n}
  \end{eqnarray*}
  where
  \[
  S_{m,n}=\sum_{k=0}^{m}\sum_{\ell=0}^{n}(-1)^{k+\ell}\,
  \frac{\Gamma(m+s+1)\Gamma(n+s+1)\Gamma(k+\ell+s)}
  {\Gamma(k+s+1)\Gamma(\ell+s+1)m!n!}\,\binom{m}{k}\binom{n}{\ell}.
  \]
  In this expression only the summand with $k=0$ does not vanish since
  \[
  \sum_{\ell=0}^{n}(-1)^{\ell}\binom{n}{\ell}\ell^{j}
  =0\quad\textrm{for }j=0,1,\ldots,n-1,
  \]
  Hence
  \begin{eqnarray*}
    S_{m,n} & = & \frac{\Gamma(m+s+1)\Gamma(n+s+1)}{\Gamma(s+1)m!n!}
    \sum_{\ell=0}^{n}(-1)^{\ell}\,\frac{\Gamma(\ell+s)}{\Gamma(\ell+s+1)}\,
    \binom{n}{\ell}\\
    & = & \frac{\Gamma(m+s+1)\Gamma(n+s+1)}{\Gamma(s+1)m!n!}\, B(s,n+1)\\
    & = & \frac{\Gamma(m+s+1)}{s\, m!}\,.
  \end{eqnarray*}
  Furthermore, $\lambda_{n}(s)-\lambda_{m}(s)=2B(n-m)$ and so
  \[
  \left\langle \varphi_{m}(0),Q(s)\varphi_{n}(0)\right\rangle
  = i\left(\frac{2}{B}\right)^{\! s}\frac{c_{m}(s)c_{n}(s)}{2B(n-m)}
  \frac{\Gamma(m+s+1)}{m!}\ .
  \]
  Now it suffices to plug in the explicit expressions for the
  normalization constants $c_{m}(s)$ and $c_{n}(s)$.
\end{proof}

Using the Stirling formula one can check the asymptotic behavior of
the matrix entries of the operator $Q(s)$ for $m$ and $n$ large.
It turns out that the operator $Q(s)$ is in some sense close to a
Hermitian operator $A(s)$ in $L^{2}(\R_{+},r\textrm{d}r)$ with the
matrix entries\begin{equation}
\left\langle \varphi_{m}(0),A(s)\varphi_{n}(0)\right\rangle =0\quad\textrm{for }m=n,\label{eq:defAs-meqn}\end{equation}
and\begin{eqnarray}
\left\langle \varphi_{m}(0),A(s)\varphi_{n}(0)\right\rangle  & = & \frac{i\sgn(s)}{2(n-m)}\,\min\left\{ \left(\frac{m+1}{n+1}\right)^{\!|s|/2},\left(\frac{n+1}{m+1}\right)^{\!|s|/2}\right\} \nonumber \\
 &  & \textrm{for }m\neq n.\label{eq:defAs-mneqn}\end{eqnarray}
Note that $A(0+)=Q(0+)$. We shall also write $Q(s)_{mn}$ instead
of\linebreak$\left\langle \varphi_{m}(0),Q(s)\varphi_{n}(0)\right\rangle $,
and similarly for $A(s)$.

\begin{lem}
  \label{thm:Qs_minus_As}
  Let $A(s)$ be the Hermitian operator in $L^{2}(\R_{+},r\textrm{d}r)$
  defined by relations (\ref{eq:defAs-meqn}) and
  (\ref{eq:defAs-mneqn}). Then $Q(s)-A(s)$ is a Hilbert-Schmidt
  operator and it holds true that
  \[
  \| Q(s)-A(s)\|_{\mathrm{HS}}\leq\frac{1}{2}\,|s|(1+|s|)^{(3+|s|)/2}.
  \]
\end{lem}

\begin{proof}
  Let us suppose for definiteness that $s>0$ and $m<n$. For $x\geq1$
  set
  \[
  g_{s}(x)=\frac{\Gamma(x+s)}{x^{s}\Gamma(x)}\,.
  \]
  One can express
  \begin{eqnarray*}
    \left|Q(s)_{mn}-A(s)_{mn}\right|
    & = & \frac{1}{2(n-m)}\left|g_{s}(m+1)^{1/2}
      -g_{s}(n+1)^{1/2}\right|\\
    &  & \times\left(\frac{m+1}{n+1}\right)^{\! s/2}g_{s}(n+1)^{-1/2}\\
    & \leq & \frac{1}{4}\, g_{s}(n+1)^{-1/2}
    \int_{0}^{1}g_{s}\!\left(m+1+(n-m)t\right)^{-1/2}\\
    &  & \qquad\times\left|g_{s}'\!\left(m+1+(n-m)t\right)\right|
    \textrm{d}t.
  \end{eqnarray*}
  Notice that
  \[
  \frac{g_{s}'(x)}{g_{s}(x)}=\frac{\Gamma'(x+s)}{\Gamma(x+s)}
  -\frac{\Gamma'(x)}{\Gamma(x)}-\frac{s}{x}\,.
  \]
  Using the well known formula for the logarithmic derivative of the
  gamma function,
  \begin{equation}
    \label{eq:logdGamma}
    -\frac{\Gamma'(z)}{\Gamma(z)}=\frac{1}{z}+\gamma+\sum_{n=1}^{\infty}
    \left(\frac{1}{n+z}-\frac{1}{n}\right),
  \end{equation}
  one finds that
  \begin{eqnarray*}
    \frac{g_{s}'(x)}{g_{s}(x)}
    & = & s\left(\sum_{n=0}^{\infty}\frac{1}{(n+x)(n+x+s)}
      -\frac{1}{x}\right)\\
    & \leq & s\left(\sum_{n=0}^{\infty}
      \frac{1}{(n+x)^{2}}-\frac{1}{x}\right)\\
    & \leq & s\left(\frac{1}{x^{2}}
      +\int_{x}^{\infty}\frac{\textrm{d}y}{y^{2}}-\frac{1}{x}\right)\\
    & = & \frac{s}{x^{2}}\,.
  \end{eqnarray*}
  Similarly,
  \begin{eqnarray*}
    \frac{g_{s}'(x)}{g_{s}(x)}
    & \geq & s\left(\int_{x}^{\infty}\frac{\textrm{d}y}{y(y+s)}
      -\frac{1}{x}\right)\\
    & = & \ln\!\left(1+\frac{s}{x}\right)-\frac{s}{x}\\
    & \geq & -\frac{s^{2}}{2x^{2}}\,.
  \end{eqnarray*}
  In particular,
  \[
  \left|g_{s}'(x)\right|\leq\frac{s(s+1)}{x^{2}}\, g_{s}(x).
  \]
  From here one derives the estimates, for $t\in[0,1]$,
  \begin{eqnarray*}
    \frac{g_{s}\!\left(m+1+(n-m)t\right)}{g_{s}(n+1)}
    & = & \exp\!\left(-\int_{m+1+(n-m)t}^{n+1}
      \frac{g_{s}'(y)}{g_{s}(y)}\,\textrm{d}y\right)\\
    & \leq & \exp\!\left(\int_{m+1}^{n+1}
      \left(\frac{s}{y}-\ln\!\left(\frac{y+s}{y}\right)\right)
      \textrm{d}y\right)\\
    & = & \exp\!\left((m+1+s)\ln\!\left(1+\frac{s}{m+1}\right)\right.\\
    &  & \left.-\,(n+1+s)\ln\!\left(1+\frac{s}{n+1}\right)\right)\\
    & \leq & (1+s)^{1+s}
  \end{eqnarray*}
  and
  \begin{eqnarray*}
    \left|Q(s)_{mn}-A(s)_{mn}\right|
    & \leq & \frac{s(s+1)}{4\, g_{s}(n+1)^{1/2}}\,
    \int_{0}^{1}\frac{g_{s}\!
      \left(m+1+(n-m)t\right)^{1/2}}{\left(m+1+(n-m)t\right)^{2}}\,
    \textrm{d}t\\
    & \leq & \frac{1}{4}\, s(1+s)^{(3+s)/2}\int_{0}^{1}
    \frac{\textrm{d}t}{\left(m+1+(n-m)t\right)^{2}}\,.
  \end{eqnarray*}
  Let $F(t)$ be a Hermitian operator in $L^{2}(\R_{+},r\textrm{d}r)$
  with the following matrix entries in the basis $\{\varphi_{n}(0)\}$:
  \[
  F(t)_{mn}=0\quad\textrm{for }m=n,
  \]
  and
  \[
  F(t)_{mn}=\left(m+1+(n-m)t\right)^{-2}\quad\textrm{for }m<n.
  \]
  Then $F(t)$ is a Hilbert-Schmidt operator and
  \begin{eqnarray*}
    \| F(t)\|_{\mathrm{HS}}^{\,2}
    & = & 2\sum_{m=0}^{\infty}\sum_{n=m+1}^{\infty}
    \left(m+1+(n-m)t\right)^{-4}\\
    & \leq & 2\sum_{m=0}^{\infty}\int_{0}^{\infty}
    \frac{\textrm{d}y}{(m+1+ty)^{4}}\\
    & = & \frac{2}{3t}\sum_{m=0}^{\infty}\frac{1}{(m+1)^{3}}\\
    & \leq & \frac{1}{t}\,.
  \end{eqnarray*}
  Hence
  \begin{eqnarray*}
    \| Q(s)-A(s)\|_{\mathrm{HS}}
    & \leq & \frac{1}{4}\, s(1+s)^{(3+s)/2}
    \int_{0}^{1}\| F(t)\|_{\mathrm{HS}}\,\textrm{d}t\\
    & \leq & \frac{1}{2}\, s(1+s)^{(3+s)/2}.
  \end{eqnarray*}
  This proves the lemma.
\end{proof}

Combining Lemma~\ref{thm:bound-As} and Lemma~\ref{thm:Qs_minus_As} we
deduce that the operator $Q(s)$ is actually bounded.

\begin{lem}
  \label{thm:estimQs}
  The operator $Q(s)$ is bounded and its norm satisfies the estimate
  \[
  \|Q(s)\|\leq\frac{\pi}{2}+12\,|s|+\frac{1}{2}\,|s|(1+|s|)^{(3+|s|)/2}.
  \]
\end{lem}

\begin{proof}
  Let $A(s)$ be the Hermitian operator in $L^{2}(\R_{+},r\textrm{d}r)$
  defined by relations (\ref{eq:defAs-meqn}) and
  (\ref{eq:defAs-mneqn}).  According to Lemma~\ref{thm:bound-As} it
  holds true that
  \[
  \|A(s)\|\leq\frac{1}{2}\left(\pi+\left(\frac{\sqrt{2}}{3}+4\right)
    \pi^{2}\frac{|s|}{2}\right).
  \]
  Lemma~\ref{thm:Qs_minus_As} leads to the estimate
  \begin{eqnarray*}
    \| Q(s)\| & \leq & \| A(s)\|+\| Q(s)-A(s)\|\\
    & \leq & \frac{1}{2}\left(\pi+\left(\frac{1}{3\sqrt{2}}+2\right)
      \pi^{2}|s|+|s|(1+|s|)^{(3+|s|)/2}\right).
  \end{eqnarray*}
  Since $\left(1+1/(6\sqrt{2})\right)\!\pi^{2}<12$ the lemma follows.
\end{proof}

\section{The meaning  of the propagator $U_\tau(s,s_0)$}
\label{sec:propagator}

As already discussed in the Introduction the natural propagator
$U_\tau(s,s_0)$ defined in (\ref{eq:UtauUAC}) is not related in the
standard way to the Hamiltonian $\tau{}H(s)$ defined in
(\ref{eq:defH}). In particular it is not clear if $U_\tau(s,s_{0})$
maps the domain $\Dom{}H(s_{0})$ into $\Dom{}H(s)$. This is why we
propose in the Appendix the notion of a propagator weakly associated
to a Hamiltonian, see Definition~\ref{def:weak_assoc}. We should like
to emphasize that this relationship is unique, i.e. at most one
propagator can be weakly associated to a Hamiltonian.

In this section we show that $U_{\tau}$ is weakly associated to $\tau
H$ and that $(s,r)\mapsto U_{\tau}(s,s_{0})\psi_{0}(r)$ satisfies the
Schrödinger equation as a distribution for all $\psi_{0}\in
L^{2}(\R_{+},r drd\varphi)$.

\begin{prop}
  \label{prop:Utau_weakassoc_H}
  The propagator $U_\tau(s,s_0)$ is weakly associated to $\tau{}H(s)$.
\end{prop}

\begin{proof}
  Relation~(\ref{eq:UtauUAC}) means that
  \begin{displaymath}
    U_\tau(s,s_0) = V(s)e^{-i\tau\Omega(s)}C(s,s_0)
    e^{i\tau\Omega(s_0)}V(s_0)^{-1}.
  \end{displaymath}
  So starting from $C(s,s_0)$ one can reach $U_\tau(s,s_0)$ by two
  consecutive unitary transformations. The propagator $C(s,s_0)$ was
  defined in (\ref{eq:defC}). It corresponds to the Hamiltonian
  $-Q_\tau(s)$ defined in (\ref{eq:defQtau}). According to
  Lemma~\ref{thm:estimQs} the function $\|Q_\tau(s)\|=\|Q(s)\|$ is
  locally bounded and thus $C(s,s_0)$ is given by the Dyson formula,
  see relation (\ref{eq:CDyson}) in Section~\ref{sec:adiabat_theorem}.
  
  First we apply Proposition~\ref{prop:weakassoc_Abound} in which we
  set $A(t)=-Q_\tau(t)$, $D=\Dom{}H(0)$, $T(t)=\exp(-i\tau\Omega(t))$
  and
  \begin{displaymath}
    X(t) = i\left(\partial_te^{-i\tau\Omega(t)}\right)
    e^{i\tau\Omega(t)} = \tau W(t).
  \end{displaymath}
  We conclude that the propagator
  $e^{-i\tau\Omega(s)}C(s,s_0)e^{i\tau\Omega(s_0)}$ is weakly
  associated to
  \begin{displaymath}
    \tau W(s)-e^{-i\tau\Omega(s)}Q_\tau(s)e^{i\tau\Omega(s)}
    = \tau W(s)-Q(s).
  \end{displaymath}
  
  Next we apply Proposition~\ref{prop:weakassoc_VinC1} in which we set
  $\tilde{H}(t)=\tau{}W(t)-Q(t)$ and
  $\tilde{U}(t,s)=e^{-i\tau\Omega(t)}C(t,s)e^{i\tau\Omega(s)}$. Recall
  further that $V(t)$ was defined in (\ref{eq:defV}). We conclude that
  $U_\tau(s,s_0)=V(s)\tilde{U}(s,s_0)V(s_0)^{-1}$ is weakly associated
  to
  \begin{displaymath}
    \tau V(s)W(s)V(s)^{-1}-V(s)Q(s)V(s)^{-1}+i\dot{V}(s)V(s)^{-1}
    = \tau H(s).
  \end{displaymath}
  The proposition is proven.
\end{proof}

In the studied model $\HH=L^{2}(\R_{+},r\textrm{d}r)$ and so
\begin{displaymath}
  \KK=L^{2}(\R,\HH,\textrm{d}s)
  = L^{2}(\R\times\R_{+},r\textrm{d}s\textrm{d}r).
\end{displaymath}
Let $\gH=\int_{\R}^{\oplus}H(s)\,\textrm{d}s$ be the direct integral
of the family of self-adjoint operators $H(s)$ which is nothing but a
multiplication operator in $\KK$. Let $K_\tau$ be the quasi-energy
operator associated to the propagator $U_\tau(s,s_0)$ (see Appendix).
According to Proposition~\ref{prop:Utau_weakassoc_H} it holds true
that
\begin{equation}
  \label{eq:Utau_weakassoc_H}
  K_\tau = \overline{-i\partial_{s}+\tau\gH}\,.
\end{equation}

To an initial condition $\psi_{0}\in\HH$ we relate the function
\newline
$\psi(s,r)=\big(U_\tau(s,0)\psi_{0}\big)(r)$ which is a locally square
integrable function in the variables $s$ and $r$. We now show that
$\psi(s,r)$ fulfills the Schrödinger equation in the space of
distributions $\sD'(\R\times\,]0,\infty[)$. Let us note that for the
proof it suffices to know that
$-i\partial_{s}+\tau\gH\subset{}K_\tau$, the stronger property
(\ref{eq:Utau_weakassoc_H}) is not necessary.

\begin{prop}
  For every $\psi_{0}\in\HH$, the function
  $\psi(s,r)=\big(U_\tau(s,0)\psi_{0}\big)(r)$ satisfies the
  Schrödinger equation in the sense of distributions.
\end{prop}

\begin{proof}
  Let $\xi\in C_{0}^{\infty}(\R\times\,]0,+\infty[)$ be an arbitrary
  real-valued test function. Set $g(s,r)=\xi(s,r)/r$. Clearly,
  $g\in\Dom(-i\partial_{s}+\tau\gH)\subset\Dom{}K_\tau$. Let
  $[a,b]\times[c,d]$ be a rectangle containing $\supp\xi$ and choose
  $\eta\in C_{0}^{\infty}(\R)$ so that $\eta\equiv1$ on a neighborhood
  of the interval $[a,b]$. From Proposition~\ref{prop:Ker0K} we know
  that $K_\tau(\eta(s)\psi(s,r))=-i\eta'(s)\psi(s,r)$. From the choice
  of $\eta$ it follows that
  \[
  0=-i\langle g,\eta'\psi\rangle_{\KK}
  =\langle g,K_\tau(\eta\psi)\rangle_{\KK}
  =\langle(-i\partial_{s}+\tau\gH)g,\eta\psi\rangle_{\KK}.
  \]
  The last term equals
  \begin{eqnarray*}
    &  & \hspace{-2em}\int_{\R\times\R_{+}}\!
    \left(i\partial_{s}\frac{1}{r}\,\xi(s,r)+\tau H(s)\frac{1}{r}\,
      \xi(s,r)\right)\!\eta(s)\psi(s,r)\, r\textrm{d}s\textrm{d}r\\
    & & \hspace{-2em}=\,\int_{\R\times\R_{+}}\!
    \left(i\partial_{s}\xi(s,r)
      +\tau\left(-\,\partial_{r}r\partial_{r}\frac{1}{r}
        +\frac{1}{r^{2}}\!\left(s+\frac{Br^{2}}{2}\right)^{\!2}\right)\!
      \xi(s,r)\right)\!\psi(s,r)\,\textrm{d}s\textrm{d}r.
  \end{eqnarray*}
  This means that
  \[
  -i\partial_{s}\psi(s,r)
  +\tau\!\left(-\frac{1}{r}\,\partial_{r}r\partial_{r}
    +\frac{1}{r^{2}}\!\left(s+\frac{Br^{2}}{2}\right)^{\!2}
  \right)\!\psi(s,r) = 0
  \]
  in the domain $\R\times\,]0,+\infty[$ in the sense of distributions.
\end{proof}

\section{Proof  of the adiabatic theorem}
\label{sec:adiabat_theorem}

We follow the strategy explained in the  Introduction. The adiabatic
propagator $U_{AD}$ (see (\ref{eq:defUAD})) and the propagator
$U_{\tau}$ defined in (\ref{eq:UtauUAC}) differ by $C$ defined by
(\ref{eq:defC}). Since
$Q_\tau(s)=e^{i\tau\Omega(s)}Q(s)e^{-i\tau\Omega(s)}$, defined in 
(\ref{eq:defQtau}), is unitarily
equivalent to $Q(s)$  it is bounded, uniformly
in $s$ on every bounded interval $[0,S]$. Hence $C(s,s_{0})$ exists
and is given by the Dyson formula:
\begin{equation}
  \label{eq:CDyson}
  C(s, s_{0})=\I+\sum_{n=1}^{\infty}i^{n}\int_{s_{0}}^{s}\textrm{d}s_{1}
  \int_{s_{0}}^{s_{1}}\textrm{d}s_{2}\ldots
  \int_{s_{0}}^{s_{n-1}}\textrm{d}s_{n}\,
  Q_{\tau}(s_{1})Q_{\tau}(s_{2})\ldots Q_{\tau}(s_{n}).
\end{equation}

The task is to estimate the norm of the integral of $Q_{\tau}$. This
will be done by the integration by parts technique developed in the
following two lemmas.

The first step is to find a bounded differentiable solution $X(s)$ of
the commutation equation
\[
Q(s)=i\,[W(s),X(s)].
\]
The operator $W(s)$ was defined in (\ref{eq:defW}). The off-diagonal
entries of the $X(s)$ are determined unambiguously,
\begin{eqnarray}
  \left\langle \varphi_{m}(0),X(s)\varphi_{n}(0)\right\rangle
  & = & -i\,\frac{\left\langle \varphi_{m}(0),Q(s)\varphi_{n}(0)
    \right\rangle }{\lambda_{m}(s)-\lambda_{n}(s)}
  \label{eq:defXmneqn}\\
  & = & -\frac{\sgn(s)}{4B(n-m)^{2}}\,\min\left\{
    \frac{\gamma_{m}(s)}{\gamma_{n}(s)},
    \frac{\gamma_{n}(s)}{\gamma_{m}(s)}\right\}
  \quad\textrm{for }m\neq n,\nonumber
\end{eqnarray}
with $\gamma_n(s)$ defined in (\ref{eq:def_gamma}). We set
\begin{equation}
  \label{eq:defXmeqn}
  \left\langle
    \varphi_{m}(0),X(s)\varphi_{n}(0)\right\rangle = 0
  \quad\textrm{for~}m=n, 
\end{equation}
and  write again $X(s)_{mn}$ instead of
$\left\langle\varphi_{m}(0),X(s)\varphi_{n}(0)\right\rangle$.

\begin{lem}
  \label{thm:estimXXdot}
  The operator $X(s)$ defined by relations (\ref{eq:defXmeqn}) and
  (\ref{eq:defXmneqn}) is bounded and its norm satisfies the
  estimate
  \[
  \| X(s)\|\leq\frac{\pi^{2}}{12\, B}\,.
  \]
  The derivative $\dot{X}(s)$ exists in the operator norm and
  satisfies the estimate
  \[
  \|\dot{X}(s)\|\leq\frac{(1+\sqrt{2})\pi^{2}}{48\, B}\,.
  \]
\end{lem}

\begin{proof}
  The operator norm of $X(s)$ is bounded from above by the
  Shur-Holmgren norm,
  \[
  \| X(s)\|\leq\|X(s)\|_{\mathrm{SH}}
  = \sup_{m\in\Z_{+}}\sum_{n=0}^{\infty}\left|X(s)_{mn}\right|
  \leq\frac{1}{2B}\sum_{k=1}^{\infty}\frac{1}{k^{2}}
  =\frac{\pi^{2}}{12\,B}\,.
  \]
  
  Suppose that $s>0$ and $m<n$. Let us estimate the derivative of
  $X(s)_{mn}$. Using (\ref{eq:logdGamma}) one finds that
  \begin{eqnarray*}
    \left(\frac{\gamma_{m}(s)}{\gamma_{n}(s)}\right)'
    & = & \frac{\gamma_{m}(s)}{2\gamma_{n}(s)}
    \left(\frac{\Gamma'(m+s+1)}{\Gamma(m+s+1)}
      -\frac{\Gamma'(n+s+1)}{\Gamma(n+s+1)}\right)\\
    & = & \frac{\gamma_{m}(s)}{2\gamma_{n}(s)}
    \sum_{k=0}^{\infty}\frac{n-m}{(k+m+s+1)(k+n+s+1)}\,.
  \end{eqnarray*}
  Hence
  \begin{eqnarray*}
    \left|\frac{\textrm{d}}{\textrm{d}s}X(s)_{mn}\right|
    & \leq & \frac{1}{8B(n-m)}\left(\frac{1}{(m+1)(n+1)}
      +\int_{1}^{\infty}\frac{\textrm{d}y}{(y+m)(y+n)}\right)\\
    & = & \frac{1}{8B(n-m)}\left(\frac{1}{(m+1)(n+1)}
      +\frac{1}{n-m}\,\ln\!\left(\frac{n+1}{m+1}\right)\right).
  \end{eqnarray*}
  Thus we get, for $m\neq n$,
  \begin{equation}
    \label{eq:estim-dXds}
    \left|\frac{\textrm{d}}{\textrm{d}s}X(s)_{mn}\right|
    \leq\frac{1}{8B}\left(\frac{1}{(m+1)(n+1)}
      +\frac{1}{|n-m|\min\{ m+1,n+1\}}\right).
  \end{equation}
  Let $\dot{X}(s)$ be a Hermitian operator in
  $L^{2}(\R_{+},r\textrm{d}r)$ with the matrix entries
  $\textrm{d}X(s)_{mn}/\textrm{d}s$. From the estimate
  (\ref{eq:estim-dXds}) we deduce that $\dot{X}(s)$ is a
  Hilbert-Schmidt operator and
  \begin{eqnarray*}
    \|\dot{X}(s)\|_{\mathrm{HS}}
    & \leq & \frac{1}{8B}\left(\sum_{m=0}^{\infty}\frac{1}{(m+1)^{2}}
      \sum_{n=0}^{\infty}\frac{1}{(n+1)^{2}}\right)^{\!1/2}\\
    &  & +\,\frac{1}{8B}\left(2\sum_{m=0}^{\infty}\frac{1}{(m+1)^{2}}
      \sum_{n=m+1}^{\infty}\frac{1}{(n-m)^{2}}\right)^{\!1/2}\\
    & = & \frac{(1+\sqrt{2})\pi^{2}}{48\, B}\,.
  \end{eqnarray*}
  Furthermore, since estimate (\ref{eq:estim-dXds}) is uniform in $s$
  one can apply the Lebesgue dominated convergence theorem to conclude
  that
  \[
  \lim_{\varepsilon\to0}\left\Vert
    \frac{1}{\varepsilon}\left(X(s+\varepsilon)-X(s)\right)
    -\dot{X}(s)\right\Vert _{\mathrm{HS}}=0.
  \]
  Hence the derivative of the operator-valued function $X(s)$ exists
  in the operator norm and equals $\dot{X}(s)$.
\end{proof}

The matrix entries of the operator $Q_{\tau}(s)$ defined in
(\ref{eq:defQtau}) equal
\[
\left\langle \varphi_{m}(0),Q_{\tau}(s)\varphi_{n}(0)\right\rangle 
=i\, e^{i\tau\left(\omega_{m}(s)-\omega_{n}(s)\right)}\left\langle
  \varphi_{m}(s),\dot{\varphi}_{n}(s)\right\rangle .
\]
Notice that the both operators $\Omega(s)$ and $W(s)=\Omega'(s)$ are
diagonal in the basis $\{\varphi_{n}(0)\}$ and therefore they commute.

\begin{lem}
  \label{thm:estim-intQu}
  It holds true that
  \[
  \Big\|\int_{0}^{s}Q_{\tau}(u)\,\textrm{d}u\Big\|
  \leq\bigg(1+\frac{1+\sqrt{2}}{8}\,|s|\bigg)\,\frac{\pi^{2}}{6B\tau}\,.
  \]
\end{lem}

\begin{proof}
  Suppose that $s>0$. The integral can be rewritten as
  follows,
  \begin{eqnarray*}
    &  & \int_{0}^{s}Q_{\tau}(u)\,\textrm{d}u
    \,=\, i\int_{0}^{s}e^{i\tau\Omega(u)}[W(u),X(u)]\,
    e^{-i\tau\Omega(u)}\,\textrm{d}u\\
    &  & \qquad\quad=\,\frac{1}{\tau}\int_{0}^{s}
    \left(\left(e^{i\tau\Omega(u)}\right)'X(u)\,
      e^{-i\tau\Omega(u)}+e^{i\tau\Omega(u)}X(u)
      \left(e^{-i\tau\Omega(u)}\right)'\right)\textrm{d}u\\
    & & \qquad\quad=\,\frac{1}{\tau}\int_{0}^{s}
    \left(\left(e^{i\tau\Omega(u)}X(u)\,
        e^{-i\tau\Omega(u)}\right)'-e^{i\tau\Omega(u)}
      \dot{X}(u)e^{-i\tau\Omega(u)}\right)\textrm{d}u\,.
  \end{eqnarray*}
  Consequently,
  \begin{eqnarray*}
    \int_{0}^{s}Q_{\tau}(u)\,\textrm{d}u
    & = & \frac{1}{\tau}\,\bigg(e^{i\tau\Omega(s)}X(s)\,
    e^{-i\tau\Omega(s)}-X(0)\\
    & & \qquad-\int_{0}^{s}e^{i\tau\Omega(u)}
    \dot{X}(u)e^{-i\tau\Omega(u)}\,\textrm{d}u\bigg).
  \end{eqnarray*}
  
  More precisely, the derivation of this equality was rather formal
  but it becomes rigorous when sandwiching the both sides with the
  scalar product
  $\left\langle\varphi_{m}(0),\cdot\,\varphi_{n}(0)\right\rangle$.
  This is to say that the both sides have the same matrix entries in
  the basis $\{\varphi_{n}(0)\}$. But since the equality concerns
  bounded operators it holds true.
  
  Using Lemma~\ref{thm:estimXXdot} one arrives at the estimate
  \begin{eqnarray*}
    \Big\|\int_{0}^{s}Q_{\tau}(u)\,\textrm{d}u\Big\|
    & \leq & \frac{1}{\tau}\left(\| X(s)\|+\| X(0)\|
      +\int_{0}^{s}\|\dot{X}(u)\|\,\textrm{d}u\right)\\
    & \leq & \frac{\pi^{2}}{B\tau}\left(\frac{1}{6}
      +\frac{1+\sqrt{2}}{48}\,s\right).
  \end{eqnarray*}
  The lemma is proven.
\end{proof}

We can now show that the adiabatic propagator $U_{AD}(s,0)$ (see
(\ref{eq:defUAD})) is close to the propagator
$U_{\tau}(s,0)=U_{AD}(s,0)C(s,0)$ defined in (\ref{eq:UtauUAC})
provided the adiabatic parameter $\tau$ is large.

\begin{prop}
  \label{prop:AdiabApprox}
  It holds true that
  \[
  \| U_{\tau}(s,0)-U_{AD}(s,0)\|\leq M(s)\,
  e^{|s|M(s)}\frac{\pi}{3B\tau}
  \]
  where
  \begin{equation}
    \label{eq:def_M}
    M(s)=\frac{\pi}{2}+12\,|s|+\frac{1}{2}\,|s|(1+|s|)^{(3+|s|)/2}.
  \end{equation}
\end{prop}

\begin{proof}
  According to Lemma~\ref{thm:estimQs}, $\|Q(s)\|\leq M(s)$, and from
  Lemma~\ref{thm:estim-intQu} one easily deduces that
  \[
  \Big\|\int_{0}^{s}Q_{\tau}(u)\,\textrm{d}u
  \Big\|\leq\frac{\pi}{3B\tau}\, M(s).
  \]
  Using formula (\ref{eq:UtauUAC}) one can estimate
  \begin{eqnarray*}
    && \hspace{-2em} 
    \| U_{\tau}(s,0)-U_{AD}(s,0)\| \,=\, \| C(s,0)-\I\|\\
    && \hspace{7em}
    \,\leq\, \sum_{n=1}^{\infty}\int_{0}^{|s|}\textrm{d}s_{1}
    \ldots\int_{0}^{s_{n-2}}\textrm{d}s_{n-1}\,\| Q_{\tau}(s_{1})\|
    \ldots\| Q_{\tau}(s_{n-1})\|\\
    && \hspace{7em}
    \qquad\times\,\Big\|\int_{0}^{s_{n-1}}\textrm{d}s_{n}\,
    Q_{\tau}(s_{n})\Big\|\\
    && \hspace{7em}
    \,\leq\, \frac{\pi}{3B\tau}\sum_{n=1}^{\infty}M(s)^{n}
    \int_{0}^{|s|}\textrm{d}s_{1}\ldots
    \int_{0}^{s_{n-2}}\textrm{d}s_{n-1}\\
    && \hspace{7em}
    \,=\, \frac{\pi}{3B\tau}\sum_{n=1}^{\infty}M(s)^{n}
    \frac{|s|^{n-1}}{(n-1)!}\,.
  \end{eqnarray*}
  The proposition is proven.
\end{proof}

\section{The general dependence on time}
\label{sec:general_dependence}

Here we show that the adiabatic theorem extends to Hamiltonians of the
form
\begin{displaymath}
  H^\zeta(s) = H\big(\zeta(s)\big)
\end{displaymath}
where $H(s)$ is defined in (\ref{eq:defH}) and $\zeta\in{}C^2(\R)$ is
a real-valued function. In order to simplify the discussion and to
avoid considering discontinuities (recall that $Q(s)$ is discontinuous
at $s=0$) we shall further assume that $\zeta'(s)>0$ and $\zeta(0)=0$.

Set
\begin{displaymath}
  V^\zeta(s) = V\big(\zeta(s)\big),\textrm{~}
  W^\zeta(s) = W\big(\zeta(s)\big),\textrm{~}
  \Omega^\zeta(s) = \int_0^s W^\zeta(u)\,\dd u.
\end{displaymath}
Let $C^\zeta(s,s_0)$ be the propagator related via the Dyson formula
to the Hamiltonian $-Q_\tau^\zeta(s)$ where
\begin{displaymath}
  Q_\tau^\zeta(s) = \exp\!\left(i\tau\Omega^\zeta(s)\right)
  Q^\zeta(s)\exp\!\left(-i\tau\Omega^\zeta(s)\right),\textrm{~}
  Q^\zeta(s) = \zeta'(s)Q\big(\zeta(s)\big).
\end{displaymath}
Exactly in the same way as in the proof of
Proposition~\ref{prop:Utau_weakassoc_H} one can show that the
propagator
\begin{displaymath}
  U_\tau^\zeta(s,s_0) = V^\zeta(s)
  \exp\!\left(-i\tau\Omega^\zeta(s)\right)C^\zeta(s,s_0)
  \exp\!\left(i\tau\Omega^\zeta(s_0)\right)V^\zeta(s_0)^{-1}
\end{displaymath}
is weakly associated to the Hamiltonian $H^\zeta(s)$. The adiabatic
propagator now reads
\begin{displaymath}
  U_{AD}^\zeta(s,s_0) = V^\zeta(s)
  \exp\!\left(-i\tau\big(\Omega^\zeta(s)-\Omega^\zeta(s_0)\big)\right)
  V^\zeta(s_0)^{-1}.
\end{displaymath}

\begin{prop}
  Assume that $\zeta\in{}C^2(\R)$, $\zeta'(s)>0$ and
  $\zeta(0)=0$. Then there exists a locally bounded function
  $m^\zeta(s)$ such that
  \begin{displaymath}
    \forall s\in\R,\textrm{~}
    \left\|U_\tau^\zeta(s,0)-U_{AD}^\zeta(s,0)\right\|
    \leq \frac{m^\zeta(s)}{B\tau}\,.
  \end{displaymath}
\end{prop}

\begin{proof}
  Suppose for definiteness that $s>0$. Recall that
  $\|Q(s)\|\leq{}M(s)$ where $M(s)$ was defined in (\ref{eq:def_M}).
  The operator-valued function
  \begin{displaymath}
    X^\zeta(s)=\zeta'(s){}X(\zeta(s)),
  \end{displaymath}
  with $X(s)$ being defined in (\ref{eq:defXmneqn}) and
  (\ref{eq:defXmeqn}), satisfies the commutation equation
  \begin{displaymath}
    Q^\zeta(s) = i\,[W^\zeta(s),X^\zeta(s)].
  \end{displaymath}
  Quite analogously as in the proof of Lemma~\ref{thm:estim-intQu} one
  derives the estimate
  \begin{displaymath}
    \left\|\int_0^s Q_\tau^\zeta(u)\,\dd u\right\|
    \leq \frac{1}{\tau}\left(\|X^\zeta(s)\|+\|X^\zeta(0)\|
      +\int_0^s \|\dot{X}^\zeta(u)\|\,\dd u\right).
  \end{displaymath}
  In virtue of Lemma~\ref{thm:estimXXdot} we have
  \begin{displaymath}
    \|X^\zeta(s)\| \leq \frac{\pi^2}{12B}\,\zeta'(s)
  \end{displaymath}
  and
  \begin{displaymath}
    \int_0^s \|\dot{X}^\zeta(u)\|\,\dd u
    \leq \frac{\pi^2}{12B}\int_0^s |\zeta''(u)|\,\dd u
    + \frac{(1+\sqrt{2})\pi^2}{48B}\int_0^s \zeta'(u)^2\,\dd u\,.
  \end{displaymath}
  Hence
  \begin{displaymath}
    \left\|\int_0^s Q_\tau^\zeta(u)\,\dd u\right\|
    \leq \frac{q^\zeta(s)}{B\tau}
  \end{displaymath}
  where
  \begin{displaymath}
    q^\zeta(s) = \frac{\pi^2}{12}\left(\zeta'(0)
      +\sup_{0\leq u\leq s}\zeta'(u)
    +\int_0^s |\zeta''(u)|\,\dd u
    + \frac{1+\sqrt{2}}{4}\int_0^s \zeta'(u)^2\,\dd u\right).
  \end{displaymath}
  Finally one can proceed similarly as in the proof of
  Proposition~\ref{prop:AdiabApprox} to derive the estimate
  \begin{displaymath}
    \left\|U_\tau^\zeta(s,0)-U_{AD}^\zeta(s,0)\right\|
    = \left\|C^\zeta(s,0)-\I\right\|
    \leq \exp\!\left(\int_0^{\zeta(s)}M(v)\,\dd v\right)
    \frac{q^\zeta(s)}{B\tau}\,.
  \end{displaymath}
  This completes the proof.
\end{proof}

\makeatletter
\renewcommand{\@seccntformat}[1]{}
\makeatother

\setcounter{section}{0}
\renewcommand{\thesection}{\Alph{section}}

\setcounter{equation}{0}
\renewcommand{\theequation}{\Alph{section}.\arabic{equation}}

\appsection{Propagator weakly associated to a Hamiltonian}

By a propagator $U(t,s)$ we mean a family of unitary operators in a
separable Hilbert space $\HH$ depending on $t,s\in\R$ which satisfies
the conditions:

\renewcommand{\labelenumi}{(\roman{enumi})} 
\begin{enumerate}
\item $U(t,s)$ is strongly continuous jointly in $t$, $s$,
\item the Chapman-Kolmogorov equality is satisfied, i.e.
  \[
  \forall t,s,r\in\R,\textrm{{ }}U(t,r)U(r,s)=U(t,s).
  \]
\end{enumerate}
Let $H(t)$, $t\in\R$, be a family of self-adjoint operators in $\HH$.
The domain may depend on $t$. The standard way how one relates a
propagator $U(t,s)$ to $H(t)$ is based on the following two
requirements:

\renewcommand{\labelenumi}{(\roman{enumi})} 

\begin{enumerate}
\item $\forall t,s\in\R,\textrm{{ }}U(t,s)\big(\Dom H(s)\big)=\Dom H(t)$,
\item $\forall\psi\in\Dom H(s),\forall t\in\R,
  \textrm{{ }}i\partial_{t}U(t,s)\psi=H(t)U(t,s)\psi$.
\end{enumerate}
Clearly, if a propagator exists then it is unique. In some situations,
however, these requirements may turn out to be unnecessarily strong.
In particular this is true for the model studied in the current paper.
The heart of the problem is illustrated on the following example.

Let $A(t)$ be a family of bounded Hermitian operators in $\HH$ which
is uniformly bounded. Then the propagator exits and is given by the
Dyson formula. Let us call it $C(t,s)$. Let $D\subset\HH$ be a dense
linear subspace, and let $T(t)$ be a strongly continuous family of
unitary operators such that $D$ is invariant with respect to $T(t)$
and for every $\psi\in D$ there exists the derivative
$\partial_{t}T(t)\psi$.  Furthermore, suppose that
$X(t)=i\dot{T}(t)T(t)^{-1}$, with $\Dom X(t)=D$, is a self-adjoint
operator for all $t$ (the dot designates the derivative).  A formal
computation gives
\[
T(t)\Big(-i\partial_{t}+A(t)\Big)T(t)^{-1}
=-i\partial_{t}+X(t)+T(t)A(t)T(t)^{-1}.
\]
If $C(t,s)$ preserved the domain $D$ then the propagator
$T(t)C(t,s)T(s)^{-1}$ would solve the Schrödinger equation for
$X(t)+T(t)A(t)T(t)^{-1}$ on $D$. Thus it is natural to associate it to
this family of self-adjoint operators. The hypothesis on $C(t,s)$ need
not be, however, satisfied since $A(t)$ is an arbitrary family of
bounded operators and so $C(t,s)$ will in general not preserve this
domain.

In this appendix we propose a way how to associate a propagator to a
given time-dependent Hamiltonian in a weak sense. This association is
more general than the standard one (which supposes a constant domain
and solving the Schrödinger equation in the strong sense) and it is
still unique (i.e.: there is at most one propagator weakly associated
to a given time dependent Hamiltonian).

Here we develop this approach only to an extent which makes it
possible to apply these ideas to the studied model with a
time-dependent Aharonov-Bohm flux.  In particular, the described
example is covered by Proposition~\ref{prop:weakassoc_Abound} below.

Let $\XX$ be a Banach space. We shall say that a vector-valued
function $f:\R\to\XX$ is absolutely continuous on $\R$ if it is
absolutely continuous on every compact interval $I\subset\R$. By the
symbol $\AC(\R,\XX)$ (or just $\AC$ if there is no danger of
misunderstanding) we shall denote the space of all absolutely
continuous vector-valued functions $f(t)$ such that the derivative
$f'(t)$ exists almost everywhere on $\R$. In such a case the function
$\| f'(t)\|$ is locally integrable and
$f(t)=f(0)+\int_{0}^{t}f'(s)\,\textrm{d}s$
\cite[Theorem~3.8.6]{HillePhillips}. If the Banach space $\XX$ has the
Radon-Nikodym property then the space $\AC(\R,\XX)$ coincides with the
space of absolutely continuous vector-valued functions $AC(\R,\XX)$.
Let us recall that $\XX$ is said to have the Radon-Nikodym property if
the fundamental theorem of calculus holds, i.e. if any absolutely
continuous function is the antiderivative of a Bochner integrable
function. For example, separable Hilbert spaces are known to have the
Radon-Nikodym property \cite{DiestelUhl}.

Clearly, if $f,g\in{}AC(\R,\HH)$ then the function $\langle
f(t),g(t)\rangle$ is absolutely continuous and
\[
\partial_{t}\langle f(t),g(t)\rangle
=\langle f'(t),g(t)\rangle+\langle f(t),g'(t)\rangle\textrm{ a.e.}
\]
Similarly, if $A\in\AC(\R,\BB(\HH))$ and $f\in{}AC(\R,\HH)$ then
$A(t)f(t)\in{}AC(\R,\HH)$ and
\[
\partial_{t}A(t)f(t)=\dot{A}(t)f(t)+A(t)f'(t)\textrm{ a.e.}
\]

Let $\{ e_{k}\}$ be an orthonormal basis in $\HH$. A vector-valued
function $f(t)=\sum\eta_{k}(t)e_{k}$ belongs to $AC(\R,\HH)$ if and
only if the following two conditions are satisfied:

\renewcommand{\labelenumi}{(\roman{enumi})} 

\begin{enumerate}
\item $\exists a\in\R$ such that $\sum_k|\eta_{k}(a)|^{2}<\infty$,
\item $\forall k,\textrm{ }\eta_{k}\in AC$, and
  $\left(\sum_k|\eta_{k}'(t)|^{2}\right)^{1/2}\in
  L_{\textrm{loc}}^{1}(\R)$.
\end{enumerate}
From here one easily derives the following criterion (alternatively,
one can again consult \cite[Theorem~3.8.6]{HillePhillips}).

\begin{alem}\label{lem:f_AC_iff}
  A vector-valued function $f:\R\to\HH$ belongs to $AC(\R,\HH)$ if and
  only if the following two conditions are satisfied:
  \renewcommand{\labelenumi}{(\roman{enumi})}
  \begin{enumerate}
  \item there exists a total set $\TT\subset\HH$ such that for all
    $\psi\in\TT$, $\langle\psi,f(t)\rangle$ is absolutely continuous,
  \item the derivative $f'(t)$ exists a.e. and
    $\|f'(t)\|\in{}L_{\mathrm{{loc}}}^{1}(\R)$.
  \end{enumerate}
\end{alem}

Set $\KK=L^{2}(\R,\HH,\textrm{d}t)$. Let us recall that to every
propagator $U(t,s)$ on $\HH$ one can relate a unique self--adjoint operator $K$ in $\KK$
which is the  generator of the one-parameter group of unitary
operators $\exp(-i\sigma K)$, $\sigma\in\R$, defined by
\[
\left(e^{-i\sigma K}f\right)\!(t)=U(t,t-\sigma)f(t-\sigma)
\]
\cite{Howland}. $K$ is called the quasi-energy operator. Equivalently,
\begin{equation}
  \label{eq:U_dert_Ustar}
  K = \gU(-i\partial_{t})\gU^{\ast}\quad
  \textrm{where }\gU=\int_{\R}^{\oplus}U(t,0)\,\textrm{d}t.
\end{equation}
So $f\in\Dom K$ if and only if
$U(t,0)^{-1}f(t)\in\Dom(-i\partial_{t})$ which means that
$f\in{}L^{2}$, $U(t,0)^{-1}f(t)\in{}AC$ and
$\big(U(t,0)^{-1}f(t)\big)'\in L^{2}$.

From (\ref{eq:U_dert_Ustar}) one concludes that the spectrum of $K$ is
purely absolutely continuous and coincides with $\R$. So the kernel of
$K$ is always trivial. It seems to be natural, however, to introduce a
generalized kernel of $K$, called $\Ker_{0}K$, as follows:
\begin{eqnarray*}
  \Ker_{0}K & = & \{ f\in L_{\textrm{loc}}^{2}(\R,\HH,\textrm{d}t);
  \textrm{ }\forall\eta\in C_{0}^{\infty}(\R),
  \textrm{ }\eta f\in\Dom K\\
  & & \textrm{ and }K(\eta f)=-i\eta'f\}.
\end{eqnarray*}
Since $K$ can be very roughly imagined as the formal operator
$-i\partial_{t}+H(t)$ the elements of $\Ker_{0}K$ can be regarded as
solutions of the Schrödinger equation in a weak sense.

\begin{aprop}
  \label{prop:Ker0K}
  Let $U(t,s)$ be a propagator and let $K$ be the quasi-energy
  operator associated to it. Then it holds
  \[
  \Ker_{0}K=\{ U(t,0)\psi;\textrm{ }\psi\in\HH\}.
  \]
\end{aprop}

\begin{proof}
  If $f(t)=U(t,0)\psi$, with $\psi\in\HH$, and
  $\eta\in{}C_{0}^{\infty}(\R)$ then, in $\KK$, there exists the
  derivative
  \[ i\,\frac{\textrm{d}}{\textrm{d}\sigma}\left(e^{-i\sigma K}\eta
    f\right)\negmedspace(t)\Big|_{\sigma=0}
  =i\,\frac{\textrm{d}}{\textrm{d}\sigma}
  \big(\eta(t-\sigma)U(t,0)\psi\big)\Big|_{\sigma=0}=-i\eta'(t)f(t).
  \]
  Hence, by the Stone theorem, $\eta f\in\Dom K$ and
  $K(\eta{}f)=-i\eta'f$.
  
  Conversely, suppose that $f\in\Ker_{0}K$ and set
  $g(t)=U(t,0)^{-1}f(t)$.  Let $\eta$ be a test function. From
  (\ref{eq:U_dert_Ustar}) one deduces that
  $\eta{}g\in\Dom(-i\partial_{t})$ and
  \[
  \partial_{t}\big(\eta(t)g(t)\big)=\eta'(t)g(t)\textrm{ a.e.}
  \]
  Since $\eta\in C_{0}^{\infty}(\R)$ is arbitrary this implies that
  $g(t)\in{}AC(\R,\HH)$ and $g'(t)=0$ a.e. Consequently,
  $g(t)=\psi\in\HH$ is a constant vector-valued function and
  $f(t)=U(t,0)\psi$.
\end{proof}

It is known that the correspondence between the propagators and the
quasi-energy operators is one-to-one
\cite[Remark (1) on p.321]{Howland}. On one hand, by the very
definition, $K$ is unambiguously determined by $U(t,s)$. On the other
hand, if $U(t,s)$ and $U_{1}(t,s)$ are two propagators with equal
quasi-energy operators, $K=K_{1}$, then $U(t,s)=U_{1}(t,s)$. This
uniqueness result is also a straightforward corollary of
Proposition~\ref{prop:Ker0K}. Actually, Proposition~\ref{prop:Ker0K}
implies that for every $\psi\in\HH$ there exists $\psi_{1}\in\HH$ such
that $U(t,0)\psi=U_{1}(t,0)\psi_{1}$ for all $t$ (we use the strong
continuity of the propagators). By setting $t=0$ one finds that
$\psi=\psi_{1}$. Hence $U(t,0)\psi=U_{1}(t,0)\psi$ for all
$\psi\in\HH$. Consequently,
\[
U(t,s)=U(t,0)U(s,0)^{-1}=U_{1}(t,0)U_{1}(s,0)^{-1}=U_{1}(t,s).
\]

For a family of self-adjoint operators $H(t)$, $t\in\R$, set
$\gH=\int_{\R}^{\oplus}H(t)\,\textrm{d}t$.  This means that $f\in\KK$
belongs to $\Dom\gH$ if and only if $f(t)\in\Dom H(t)$ a.e. and
$\|H(t)f(t)\|\in L^{2}(\R,\textrm{d}t)$. Then $\gH$ is a self-adjoint
operator in $\KK$. In what follows we shall always suppose that the
intersection $\Dom(-i\partial_{t})\cap\Dom\gH$ is dense in $\KK$. For
example, this is true in the case when the domain $\Dom H(t)$ is
independent of $t$. Consequently, $-i\partial_{t}+\gH$ is a densely
defined symmetric operator.

\begin{adef}
  \label{def:weak_assoc}
  We shall say that a propagator $U(t,s)$ is \emph{weakly associated}
  to $H(t)$ if
  \begin{equation}
    K=\overline{-i\partial_{t}+\gH}\,.\label{eq:weak_assoc}
  \end{equation}
\end{adef}

Notice that equality (\ref{eq:weak_assoc}) is equivalent to the following
two conditions:

\renewcommand{\labelenumi}{(\roman{enumi})} 

\begin{enumerate}
\item $-i\partial_{t}+\gH\subset K$,
\item $-i\partial_{t}+\gH$ is essentially self-adjoint.
\end{enumerate}
Furthermore, it is important to note that this definition still guarantees
the uniqueness, i.e. to $H(t)$ one can weakly associate at most one
propagator $U(t,s)$. Actually, if $U(t,s)$ and $U_{1}(t,s)$ are
weakly associated to $H(t)$ then $K=K_{1}$ according to equality
(\ref{eq:weak_assoc}). But due to the one-to-one correspondence between
the propagators and the quasi-energy operators we have $U(t,s)=U_{1}(t,s)$.

Now we are ready to formulate and prove two propositions which are
directly applicable to the model studied in this paper.

\begin{aprop}
  \label{prop:weakassoc_Abound}
  Let $A(t)$ be a family of bounded self--adjoint operators in $\HH$ which
  is locally bounded. Let $C(t,s)$ be the propagator associated to
  $A(t)$ via the Dyson formula. Let $D\subset\HH$ be a dense linear
  subspace and let $T(t)$ be a strongly continuous family of unitary
  operators in $\HH$ obeying the conditions:
  \renewcommand{\labelenumi}{(\roman{enumi})}
  \begin{enumerate}
  \item $\forall t\in\R,\textrm{{ }}T(t)D=D$,
  \item $\forall\psi\in D$, $T(t)\psi$ is continuously differentiable,
  \item $\forall t\in\R$, $X(t)=i\dot{T}(t)T(t)^{-1}$, with
    $\Dom{}X(t)=D$, is a self-adjoint operator.
  \end{enumerate}
  Then the propagator $T(t)C(t,s)T(s)^{-1}$ is weakly associated to
  the family $X(t)+T(t)A(t)T(t)^{-1}$.
\end{aprop}

\begin{proof}
  Set
  \[
  Y(t)=X(t)+T(t)A(t)T(t)^{-1},\textrm{{ }}\gY
  =\int_{\R}^{\oplus}Y(t)\,\textrm{d}t,\textrm{{ }}\gT
  =\int_{\R}^{\oplus}T(t)\,\textrm{d}t.
  \]
  Let $K_{Y}$ be the quasi-energy operator associated to the
  propagator $T(t)C(t,s)T(s)^{-1}$. Set
  \[
  C(t)=C(t,0),\textrm{{~}}\gC=\int_{\R}^{\oplus}C(t)\,\textrm{d}t.
  \]
  $C(t)$ is a family of unitary operators which satisfies
  $C(t)\in\AC(\R,\BB(\HH))$ and $A(t)=i\dot{C}(t)C(t)^{-1}$.
  
  (i) Let us verify that
  \[
  -i\partial_{t}+\gY\subset K_{Y}
  =\gT\gC(-i\partial_{t})\gC^{-1}\gT^{-1}.
  \]
  Suppose that a vector-valued function $f:\R\to\HH$ belongs to
  $\Dom(-i\partial_{t}+\gY)$.  This happens if and only if $f$ obeys
  the conditions: $f\in{}L^{2}$, $f\in{}AC$, $f'\in{}L^{2}$,
  $f(t)\in{}D$ a.e. and $Y(t)f(t)\in L^{2}$.  In that case the
  function $T(t)^{-1}f(t)$ is differentiable a.e.  and the derivative
  \[
  \left(T(t)^{-1}f(t)\right)'=T(t)^{-1}\big(f'(t)+iX(t)f(t)\big)
  \]
  is square integrable. Moreover, if $\psi\in D$ then the function
  $\langle\psi,T(t)^{-1}f(t)\rangle=\langle T(t)\psi,f(t)\rangle$ is
  absolutely continuous. According to Lemma~\ref{lem:f_AC_iff} this
  implies that $T(t)^{-1}f(t)\in{}AC(\R,\HH)$ and consequently
  $C(t)^{-1}T(t)^{-1}f(t)\in{}AC$ as well. Furthermore, a
  straightforward computation yields
  \begin{eqnarray*}
    Y(t)f(t) & = & i\left(\dot{T}(t)T(t)^{-1}f(t)
      +T(t)\dot{C}(t)C(t)^{-1}T(t)^{-1}f(t)\right)\\
    & = & i\big(T(t)C(t)\big)'C(t)^{-1}T(t)^{-1}f(t)\\
    & = & if'(t)-iT(t)C(t)\left(C(t)^{-1}T(t)^{-1}f(t)\right)'.
  \end{eqnarray*}
  Hence $\left(C(t)^{-1}T(t)^{-1}f(t)\right)'\in L^{2}$,
  $f\in\Dom{}K_{Y}$ and $-if'(t)+Y(t)f(t)=K_{Y}f(t)$.
  
  (ii) Let us verify that $-i\partial_{t}+\gY$ is essentially
  self-adjoint.  Suppose that $g\in\Dom(-i\partial_{t}+\gY)^{\ast}$
  satisfies $(-i\partial_{t}+\gY)^{\ast}g=zg$ with $\Im(z)\neq0$. This
  means that
  \[
  \forall f\in\Dom(-i\partial_{t}+\gY),\textrm{~}
  \langle(-i\partial_{t}+\gY)f,g\rangle_{\KK}=z\langle
  f,g\rangle_{\KK}.
  \]
  Choose $f(t)=\eta(t)T(t)\psi$ where $\psi\in D$ and
  $\eta\in{}C_{0}^{\infty}(\R)$ is real-valued. Then
  $f\in\Dom(-i\partial_{t}+\gY)$ and an easy computation shows that
  \[
  (-i\partial_{t}+\gY)f(t)=-i\eta'(t)T(t)\psi+\eta(t)T(t)A(t)\psi.
  \]
  Hence for all $\eta\in C_{0}^{\infty}(\R)$ we have
  \[
  \int_{\R}\left(i\eta'(t)\langle T(t)\psi,g(t)\rangle
    +\eta(t)\langle T(t)A(t)\psi,g(t)\rangle\right)\textrm{d}t
  =z\int_{\R}\eta(t)\langle T(t)\psi,g(t)\rangle\textrm{d}t\,.
  \]
  Setting
  \[
  F(t)=\langle T(t)\psi,g(t)\rangle,\textrm{ }G(t)
  =\langle T(t)A(t)\psi,g(t)\rangle,
  \]
  we find that
  \begin{equation}
    \label{eq:der_Ft}
    -i\partial_{t}F(t)+G(t)=zF(t)
  \end{equation}
  in the sense of distributions. Since both $F(t)$ and $G(t)$ are
  locally integrable, a standard result from the theory of
  distributions tells us that $F(t)$ is absolutely continuous and
  equality (\ref{eq:der_Ft}) holds true in the usual sense. Moreover,
  equality (\ref{eq:der_Ft}) implies that
  \[
  \partial_{t}\left(e^{2\Im(z)t}|F(t)|^{2}\right)
  =2\,e^{2\Im(z)t}\Im\!\big(\overline{F(t)}G(t)\big).
  \]
  
  Let us now choose an orthonormal basis $\{\psi_{k}\}$ whose elements
  all belong to the domain $D$. Let us write $F_{k}$ instead of $F$
  and $G_{k}$ instead of $G$ when replacing $\psi$ by $\psi_{k}$. We
  have derived the equality
  \begin{equation}
    \label{eq:Fk_eq_int}
    |F_{k}(t)|^{2}=e^{-2\Im(z)(t-a)}|F_{k}(a)|^{2}
    +2\int_{a}^{t}e^{-2\Im(z)(t-s)}
    \Im\!\big(\overline{F_{k}(s)}G_{k}(s)\big)\textrm{d}s
  \end{equation}
  which is valid for all $k$ and all $a,t\in\R$. Observe
  that
  \begin{eqnarray*}
    &  & \sum_{k}|F_{k}(t)|^{2}=\| g(t)\|^{2}\textrm{ a.e.},\\
    & & \sum_{k}|F_{k}(s)||G_{k}(s)|\leq\| g(s)\|\,\|
    A(s)T(s)^{-1}g(s)\|\in
    L_{\textrm{loc}}^{1}(\R,\textrm{d}s)\textrm{ a.e.},
  \end{eqnarray*}
  and
  \[
  \sum_{k}\overline{F_{k}(s)}G_{k}(s)
  =\left\langle g(s),T(s)A(s)T(s)^{-1}g(s)\right\rangle \in\R
  \textrm{{ }a.e.}
  \]
  Summing in $k$ in equality (\ref{eq:Fk_eq_int}) we find that
  \[
  \|g(t)\|=e^{-\Im(z)(t-a)}\| g(a)\|
  \]
  for almost all $a,t\in\R$. Since $\| g(t)\|$ is square integrable
  this is possible only if $g(t)=0$ a.e.
\end{proof}

Proposition~\ref{prop:weakassoc_Abound} has a corollary justifying the
adverb ``weakly'' in Definition~\ref{def:weak_assoc}.

\begin{acor}
  \label{rem:weakassoc_Abound}
  Assume that a propagator $U(t,t_{0})$ is associated as a strong
  solution of the Schrödinger equation to a time-dependent Hamiltonian
  $H(t)$ which has, however, a time-independent domain (i.e. the
  relationship between the propagator and the Hamiltonian is the usual
  one). Then $U(t,t_{0})$ is weakly associated to $H(t)$.
\end{acor}

\begin{proof}
  In Proposition~\ref{prop:weakassoc_Abound} it suffices to set
  $D=\Dom{}H(0)$, $T(t)=U(t,0)$ and $A(t)=0$. Then $X(t)=H(t)$,
  $C(t,s)=\I$ and $T(t)C(t,s)T(s)^{-1}=U(t,s)$.
\end{proof}

\begin{aprop}
  \label{prop:weakassoc_VinC1}
  Suppose that $V(t)$, $t\in\R$, is a family of unitary operators
  which is continuously differentiable in the strong sense. Let
  $\tilde{H}(t)$, $t\in\R$, be a family of self-adjoint operators such
  that $\Dom\tilde{H}(t)=D$ for all $t\in\R$. Set
  \[
  H(t)=V(t)\tilde{H}(t)V(t)^{-1}+i\dot{V}(t)V(t)^{-1}.
  \]
  If the propagator $\tilde{U}(t,s)$ is weakly associated to
  $\tilde{H}(t)$ then the propagator
  $U(t,s)=V(t)\tilde{U}(t,s)V(s)^{-1}$ is weakly associated to $H(t)$.
\end{aprop}

\begin{proof}
  Set
  \[
  \tilde{U}(t)=\tilde{U}(t,0),\textrm{ }\tilde{\gU}
  =\int_{\R}^{\oplus}\tilde{U}(t)\,\textrm{d}t,\textrm{ }\gV
  =\int_{\R}^{\oplus}V(t)\,\textrm{d}t\,.
  \]
  By the assumption,
  $\tilde{\gU}(-i\partial_{t})\tilde{\gU}^{-1}
  =\overline{-i\partial_{t}+\tilde{\gH}}$.
  We have to show that
 \[
  \gV\tilde{\gU}(-i\partial_{t})\tilde{\gU}^{-1}\gV^{-1}
  =\overline{-i\partial_{t}+\gH}\,.
  \]
  Since
  \[
  \gV\tilde{\gU}(-i\partial_{t})\tilde{\gU}^{-1}\gV^{-1}
  =\gV\overline{(-i\partial_{t}+\tilde{\gH})}\gV^{-1}
  =\overline{\gV(-i\partial_{t}+\tilde{\gH})\gV^{-1}}
  \]
  it is sufficient to verify that
  \begin{displaymath}
    \gV(-i\partial_{t}+\tilde{\gH})\gV^{-1}=-i\partial_{t}+\gH.
  \end{displaymath}
  This would also imply that $\Dom(-i\partial_{t})\cap\Dom(\gH)$ is
  dense in $\KK$.
  
  A vector-valued function $f:\R\to\HH$ belongs to
  $\Dom\!\big(\gV(-i\partial_{t}+\tilde{\gH})\gV^{-1}\big)$ if and
  only if it satisfies the conditions: $f\in L^{2}$,
  $V(t)^{-1}f(t)\in{}AC$, $\left(V(t)^{-1}f(t)\right)'\in L^{2}$,
  $V(t)^{-1}f(t)\in D$ a.e.  and $\tilde{H}(t)V(t)^{-1}f(t)\in L^{2}$.
  Let us note that from the continuous differentiability of $V(t)$ in
  the strong sense and from the uniform boundedness principle it
  follows that $\dot{V}(t)$, $t\in\R$, is a family of bounded
  operators which is locally bounded. Furthermore,
  $V(t)^\ast=V(t)^{-1}$ is continuously differentiable in the strong
  sense as well and $V(t)^{-1}\psi\in{}AC$ for all $\psi\in\HH$.
  Suppose that $f\in{}L^2$. If $V(t)^{-1}f(t)\in{}AC$ then $f'(t)$
  exists a.e. and $\|f'(t)\|$ is locally integrable, the function
  $\langle\psi,f(t)\rangle=\langle{}V(t)^{-1}\psi,V(t)^{-1}f(t)\rangle$
  is absolutely continuous for all $\psi\in\HH$ and therefore, by
  Lemma~\ref{lem:f_AC_iff}, $f(t)\in{}AC$. Similarly, the converse is
  also true. If $f(t)\in{}AC$ then $V(t)^{-1}f(t)\in{}AC$.
  
  Using these facts and the relation between $\tilde{H}(t)$ and $H(t)$
  (including that $\Dom H(t)=V(t)D$) one easily finds that the domains
  of $\gV(-i\partial_{t}+\tilde{\gH})\gV^{-1}$ and
  $-i\partial_{t}+\gH$ coincide and that
  \[
  V(t)\big(-i\partial_{t}+\tilde{H}(t)\big)V(t)^{-1}f(t)=-if'(t)+H(t)f(t)
  \]
  for every $f\in\Dom(-i\partial_{t}+\gH)$.
\end{proof}

\begin{rem*}
  Proposition~\ref{prop:weakassoc_VinC1} can be easily extended to the
  case when the family of unitary operators $V(t)$ is continuous and
  piece-wise continuously differentiable in the strong sense and in
  each point of discontinuity there exist the limits of the derivative
  both from the left and from the right.
\end{rem*}

\section*{Acknowledgements}
P.~\v{S}. wishes to acknowledge gratefully the support from the grant
No. 201/05/0857 of Grant Agency of the Czech Republic. J.~A. thanks Czech Technical University and I.~H. Toulon University  for support and hospitality. This paper is the preprint CPT-05/P.008.

\end{document}